\documentclass[traditabstract]{aa} 
\pdfoutput=1                                   
\usepackage{graphicx}
\usepackage[]{natbib}
\usepackage{pdflscape}
\usepackage{longtable}
\usepackage[varg]{txfonts}
\usepackage{enumerate}
\begin{document}
%

\title{Milky Way Demographics with the VVV Survey II.
      \thanks{Based on observations taken within the ESO
VISTA Public Survey VVV, Programme ID 179.B-2002. }}

  \subtitle{Color Transformations and  Near-Infrared Photometry 
 for 136 Million Stars in the Southern Galactic Disk
 }

   \author{M.Soto\inst{1},
          R. Barb\'a\inst{1,2},
          G. Gunthardt \inst{1,3},
          D. Minniti \inst{4,5,6,7,8,9},
          P. Lucas \inst{10}, 
          D.Majaess \inst{11},
          M. Irwin \inst{12},
          J.P. Emerson \inst{13}
          E.Gonzalez-Solares \inst{12},
          M. Hempel \inst{4,8},
          R.K. Saito \inst{4,8,14}
          S. Gurovich \inst{3,15},
          A. Roman-Lopes \inst{1},
          C. Moni-Bidin \inst{16,17}
          M.V. Santucho \inst{3},
          J.Borissova \inst{8,14},
          R. Kurtev \inst{14},
          I. Toledo \inst{4,18},
          D. Geisler \inst{16},
          M. Dominguez \inst{3,15},
          J.C. Beamin \inst{4}.
          }

   \institute{Departamento de F\'{\i}sica, Universidad de La Serena,
            Benavente 980, La Serena, Chile 
            \email{msoto@dfuls.cl}
            \and
             Instituto de Ciencias Astron\'omicas, del la Tierra y del Espacio
             (ICATE-CONICET), Av. Espa\~na Sur 1512, J5402DSP San Juan,
             Argentina
            \and
             Observatorio Astron\'omico de C\'ordoba, Universidad
             Nacional de C\'ordoba, Argentina 
            \and
             Departamento de Astronom\'{\i}a y Astrof\'{\i}sica,
             Pontificia Universidad Cat\'olica de Chile, Vicu\~na Mackena
             4860, Casilla 306, Santiago 22, Chile
             \and
             Vatican Observatory, Vatican City State V-00120, Italy
             \and
             European Southern Observatory, Vitacura 3107, Santiago,
             Chile
             \and
             Department of Astrophysical Sciences, Princeton University, Princeton NJ 08544-1001, USA
             \and
             The Milky Way Millennium Nucleus, Av. Vicu\~na Mackenna
             4860, 782-0436 Macul, Santiago, Chile
             \and
             Departamento de Ciencia Fisicas, Universidad Andres Bello, Avda. Republica 252, Santiago, Chile
             \and
             Centre for Astrophysics Research, Science and Technology Research Institute, University of Hertfordshire, Hatfield AL10 9AB, UK
             \and
             Department of Astronomy and Physics, Saint Mary's
             University, Halifax, Nova Scotia, B3K 5L3, Canada
             \and
             Institute of Astronomy, University of Cambridge, Madingley Road, Cambridge CB3 0HA, UK
             \and
             Astronomy Unit, School of Physics and Astronomy, Queen
             Mary University of London, Mile End Road, London, E1 4NS, UK
             \and
             Departamento de F\'{\i}sica y Astronom\'{\i}a,
             Universidad de Valpara\'{\i}so, Av. Gran Breta\~na 1111,
             Playa Ancha, Casilla 5030, Chile
             \and
             Instituto de Astronom\'{\i}a Te\'orica y Experimental, CONICET, Laprida 922, 5000 C\'ordoba, Argentina
             \and
             Departmento de Astronom\'{\i}a, Universidad de Concepci\'on, Casilla 160-C, Concepci\'on, Chile
             \and
             Instituto de Astronom\'{\i}a, Universidad Cat\'olica del Norte, Av. Angamos 0610, Antofagasta, Chile
             \and
             Atacama Large Millimeter Array, Alonso de C\'ordova 3107, Vitacura, Santiago, Chile
             }

   \date{Received ; accepted }

 
  \abstract
{The new multi-epoch near-infrared $VVV$  survey (VISTA Variables in
  the V\'{\i}a L\'actea) is sampling $562\ deg^2$ of the Galactic
  bulge and adjacent regions of the disk.  
  Accurate astrometry
  established for the region surveyed allows the $VVV$ data to be
  merged with overlapping surveys (e.g., GLIMPSE, WISE, 2MASS, etc.), thereby enabling
  the construction of longer baseline spectral energy distributions
  for astronomical targets.  However, in order to maximize use of the
  $VVV$ data, a set of transformation equations are required to place
  the $VVV$ $JHK_s$ photometry onto the 2MASS system.  The impetus for
  this work is to develop those transformations via a comparison of
 2MASS targets in 152 $VVV$ fields sampling the Galactic disk.  The transformation coefficients derived exhibit a reliance on variables such as extinction.   The transformed data were subsequently employed to establish a mean reddening law of $E_{J-H}/E_{H-K_s}=2.13\pm0.04$, which is the most precise determination to date and merely emphasizes the pertinence of the $VVV$ data for determining such important parameters.}

   \keywords{Galaxy: disk --
                Galaxy: stellar content --
                Galaxy: structure --
                infrared: stars --
                surveys
               }

   \titlerunning{VVV-2MASS Transformations for the Galactic Disk}
   \authorrunning{Soto et al.}  
   \maketitle
%

\section{Introduction}
The fields of the Southern Galactic disk are complicated regions to
research.   Near the Galactic plane the ISM is rich, complex, and
dust extinction is extreme and inhomogeneous at small scales.  Moreover, the surface
density of sources reaches a maximum in the Galactic plane.  Fields exhibiting
one star brighter than $K_s=18$ are catalogued every few square
arcseconds.  Owing to
the aforementioned factors, existing optical and low-spatial
resolution surveys have thus been inefficient at characterizing
populations in the Southern Galatic disk.  Hence the importance of the
VVV survey (Minniti et al. 2010; Saito et al. 2012), which is a
near-IR ESO public survey that is sampling 562 deg$^2$ of the Galactic
bulge and adjacent regions of the disk.  The survey is being carried
out via the VISTA telescope, and images are being acquired through 5
broadband filters.
The VVV fields examined here overlap with the GLIMPSE survey, which 
acquired images at 3.6, 4.5, 5.8 and 8 $\mu m$.  Thus the sources 
surveyed will have multiband photometry ranging from the near to mid-infrared.  
The region sampled is of particular interest for interstellar medium
(ISM) studies because the 4th Galactic quadrant hosts the \object{Coal Sack}, in tandem with several prominent
nebulae and areas exhibiting large star formation rates (SFR).  The inner disk region includes large numbers of open clusters
  (Borissova et al. 2011, Bica et al. 2008; Kharchenko et al. 2005 and
  references therein) and associations, 
 which allows for detailed stellar population studies.    The near-infrared nature of the surveys is particularly pertinent for such analyses since such photometry is less sensitive to dust obscuration than optical observations, thus permitting greater penetration into the
disk.  The line of sight depth in the 1st and 4th Galactic quadrants is large, and nearby foreground dwarfs stars are mixed with distant red
giants along a given line of sight.  In addition, the region sampled is pertinent for Galactic structure studies,
as there is presently no consensus on the number or delineation
of the Galaxy's major spiral arms (Benjamin et al. 2008,
Majaess et al. 2009), and the tail end of the long bar is
undercharacterized (Fig.~7 in Majaess 2010). 
\begin{figure*}[!t]
\centering
\includegraphics[width=15cm]{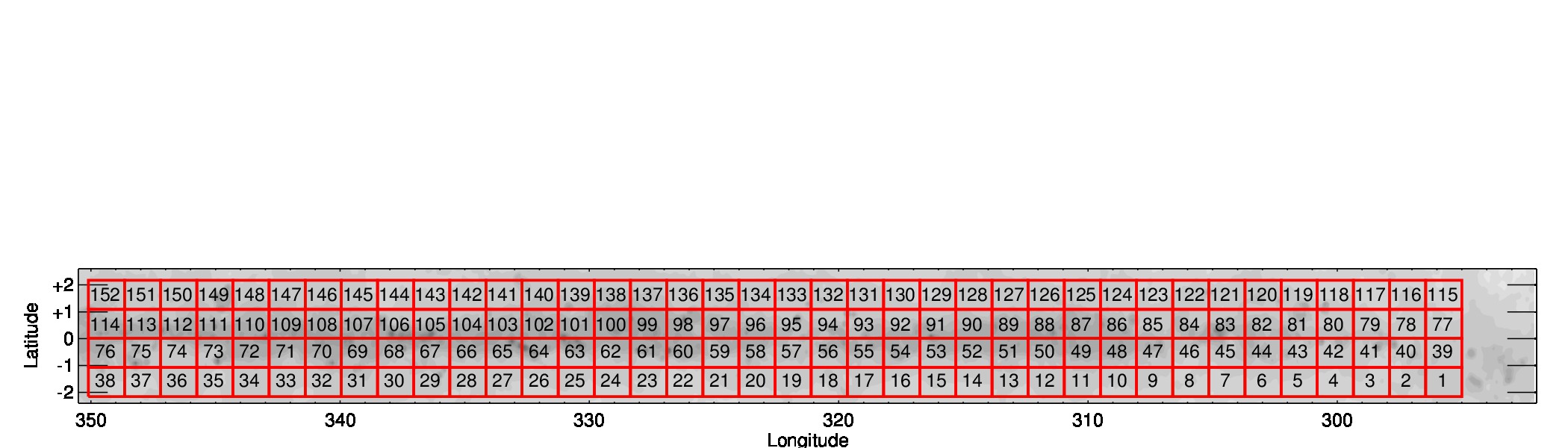}
\caption{Galactic disk fields imaged by the VVV survey.
 The VVV disk tile names start with \textbf{d}, followed by the numbering as shown in the figure.
}
\label{fig:fields1}%
\end{figure*}


Prior to VVV the 2MASS survey (Skrutskie et al. 2006) has been the
main near-infrared (near-IR) photometric survey covering much of the galactic plane,
and many models and data in the 2MASS photometric system have appeared
in the literature. In comparing with this previous work it is
therefore useful, at least until sufficient modeling is available in
the VISTA photometric system, to know the 2MASS equivalent magnitudes
to the actual VISTA magnitudes.  
In addition, such transformations allow to combine the information of
both surveys to reach a wider magnitude range (e.g. saturated stars in
the VVV catalogues can be complemented with the 2MASS magnitudes once
they are transformed to the 2MASS system).
However we note that, for the greatest accuracy, transformations between different photometric systems should in general be avoided as they will always be dependent on the (usually unknown) spectrum of the objects, but pending availability of more models in the VISTA system we find that it is expedient to estimate what VISTA magnitudes would be in the 2MASS system.

\begin{figure}
   \centering
  \includegraphics[width=9.0cm]{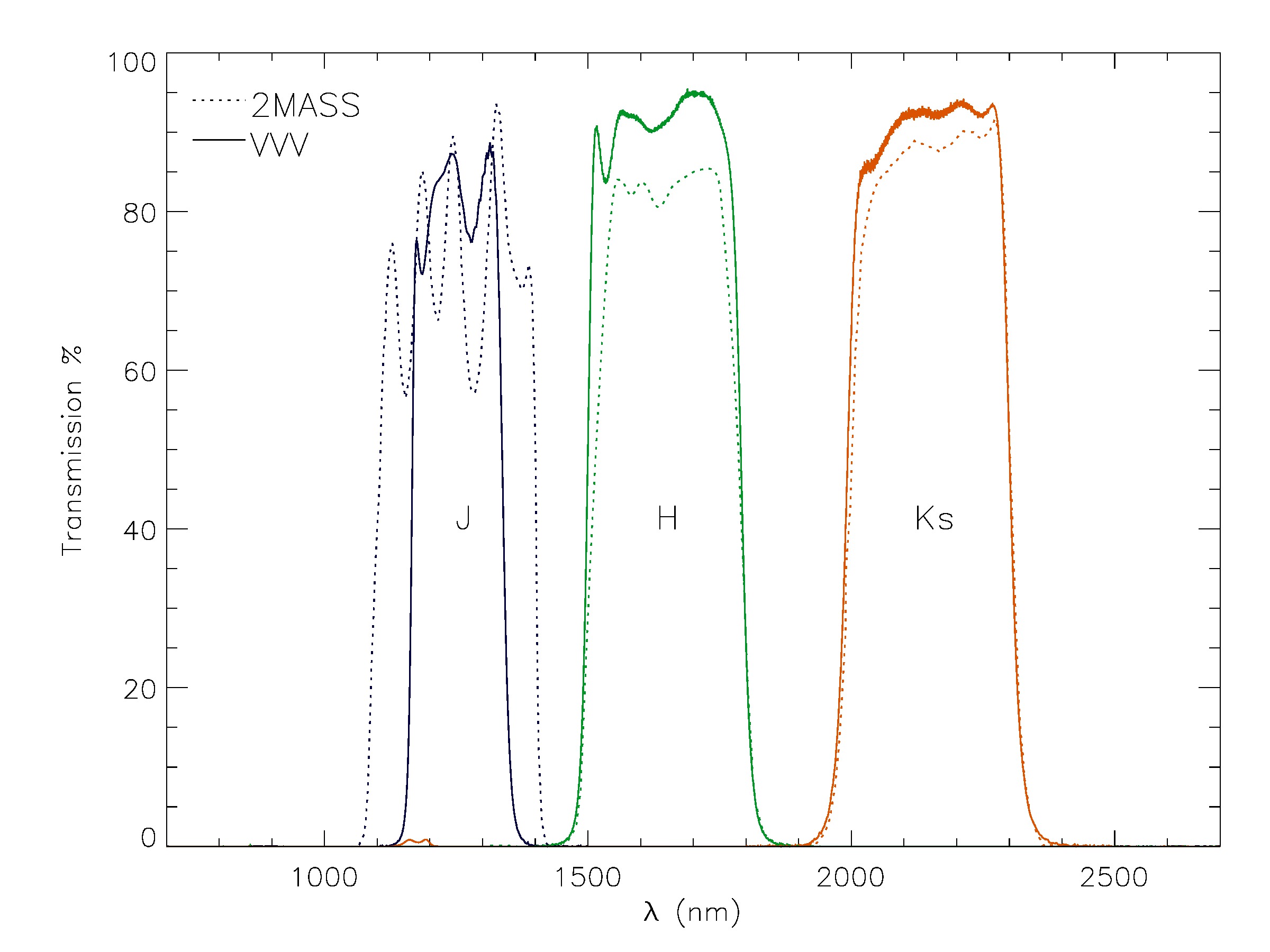}
   \caption{Transmission Curves for the 2MASS and VISTA photometric systems.}
              \label{fig:filters}%
 \end{figure}

Carpenter et al. (2001) produced transformation equations to convert
colors and magnitudes from AAO, ARNICA, CIT, DENIS, ESO, LCO, MSSSO,
SAAO and UKIRT photometric systems to 2MASS. This paper follows a similar procedure to obtain transformations linking the VVV and 2MASS systems.  A well defined set of transformation equations should be valid over a large color baseline
(i.e. $-0.5 \leq (J-K_S) \leq 4.0$) owing to the high (and strongly varying) reddening values, and the presence of
(bluer) dwarfs and (redder) giants.  In this work we present empirical $JHK_s$ color transformations for 152 tiles
 completed during year 1 (2010), photometric catalogue version 1.1
 (see observation schedule in Minniti et al. 2010), between the VISTA
 and 2MASS photometric systems.  The calibrations were inferred from
 VVV sources in the Southern Galactic disk in the region bounded by
 $-65.3^\circ \lesssim  l \lesssim -10.0^\circ$ and $-2.25^\circ
 \lesssim b \lesssim +2.25^\circ$, and apply to all the VVV
 photometry derived from the Cambridge Astronomical Survey Unit (CASU) catalogues in this area. 


\begin{figure*}
   \centering
  \includegraphics[width=8.0cm]{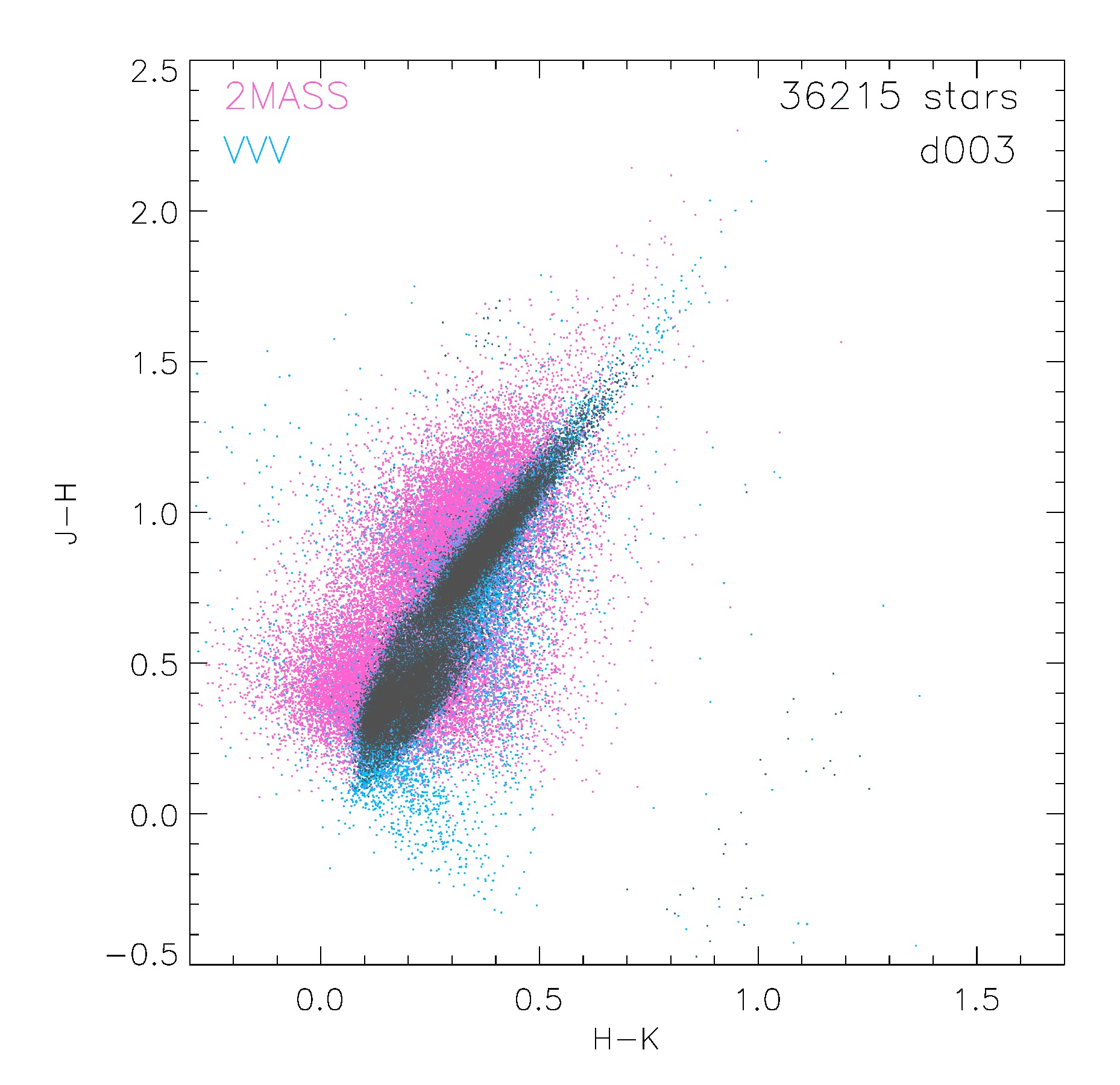}
  \includegraphics[width=8.0cm]{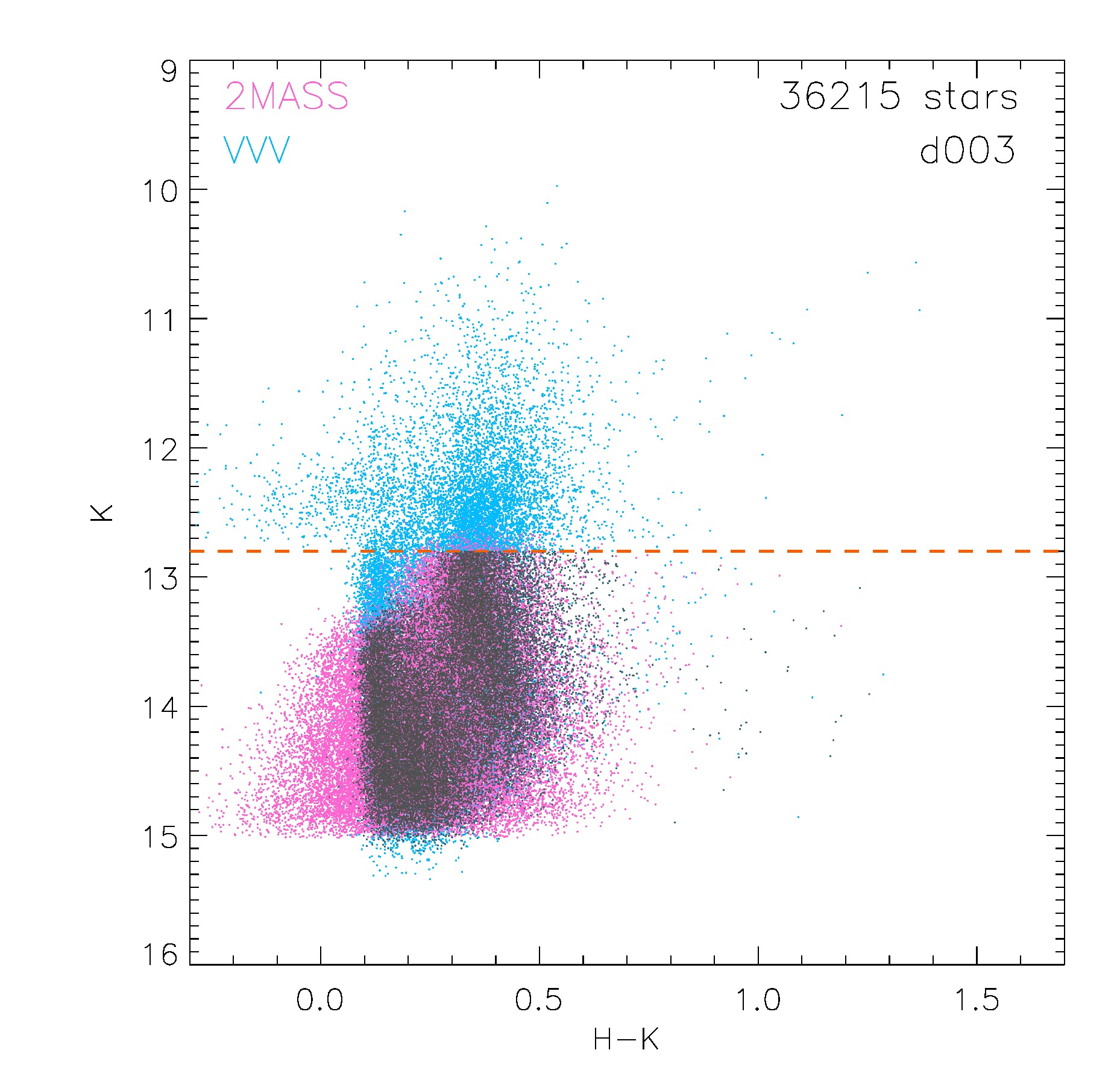}
   \caption{Color-color and color-magnitude diagrams for the tile
     \emph{d003}. VVV data (blue dots) have been matched with 2MASS
     data on the same field (pink dots), where a subsample has been
     chosen to calculate the coefficients (dark grey).
     A limiting magnitude (orange dashed line) in each band has been used to avoid
     saturated VVV stars.}
              \label{fig:cmd0}%
    \end{figure*}

This paper is organized as follows: In Section 2 a brief
overview is provided of the VVV observations and the CASU pipeline, which produces
the photometric catalogues. 
Section 3  explains the selection
procedure for the subsample of VVV-2MASS stars used to derive the photometric 
transformations. Finally, Section 4 contains the discussion of the transformation coefficients
obtained, while our conclusions are summarized in Section 5.


\section{Observations}

\subsection{VVV Observations}
Near-IR VVV observations were acquired via VISTA (Visible and
Infrared Survey Telescope for Astronomy), which is stationed at the Paranal Observatory.
 VISTA is a 4m telescope
equipped with the VIRCAM instrument (VISTA InfraRed CAMera; Emerson et
al. 2006). 
Each VVV field (called a \emph{tile}) covers 1.64 deg$^2$.  
There are 152 tiles covering about 250 deg$^2$ of the Galactic disk (Fig. \ref{fig:fields1}). The VVV tiles exhibit overlap between consecutive blocks, and for the complete survey the overlap sums to $\sim42$ deg$^2$.  The equatorial and Galactic coordinates for the center of each tile are listed in Table A.1 of Saito et al. (2012).
The VISTA IR mosaic camera has a 1.65 $deg$ diameter field of view, that is 
sampled with 16 Raytheon $2048\times2048$ arrays 
(Dalton et al. 2006). The detectors have $0.339"$ pixels which
produce a $0.6\ deg^2$ field per pointing. Each
pointing is called a \emph{"pawprint"}, 
with spacings of
$42\%$ and $90\%$  between the detectors along the $X$ and $Y$ axes,
respectively, where 6 overlapping pawprints are used to build a tile.
 
The VVV survey includes observations of the complete survey area in the 5
available filters, i.e. $Z$, $Y$, $J$, $H$, and $K_{\rm s}$. As described by Minniti
et al (2010), these multi-band observations were scheduled to be carried out during the
first year of the survey (2010), but were partly carried over into 2011 owing to various factors. The principal part
of the survey (year 2-5) will be a $K_{\rm s}$-band variability study.

\subsection{CASU pipeline: photometry}
The reduction of IR data is far more complex than reducing optical data.
IR detectors are more unstable than optical detectors and
sky emission can be several magnitudes brighter than
many IR stellar sources (Lewis et al. 2010), and marginally
fainter than the saturation limit. In the case of VVV,
the limiting magnitude for the aperture photometry of the catalogues
appears at $K_{\rm s}$=18 mag in most fields in the disk, with an expected
sky brightness at the VISTA site of $K_{\rm s} \approx$13.0 mag (Cuby et al.
2000). Moreover, IR sky emission varies over short timescales,
and changes in spatial scale can be large or small
depending on the instrument. Consequently, short exposures are needed, which subsequently increases the amount of information acquired
each night. Thus, surveys like VVV require automated
pipelines to process the large volumes of nightly data. The 
VISTA Data Flow System pipeline running at CASU handles the data processing\footnote{for more details, see
  http://casu.ast.cam.ac.uk/surveys-projects/vista/technical/photometric-properties}.


   \begin{figure*}
   \centering
  \includegraphics[width=8.0cm]{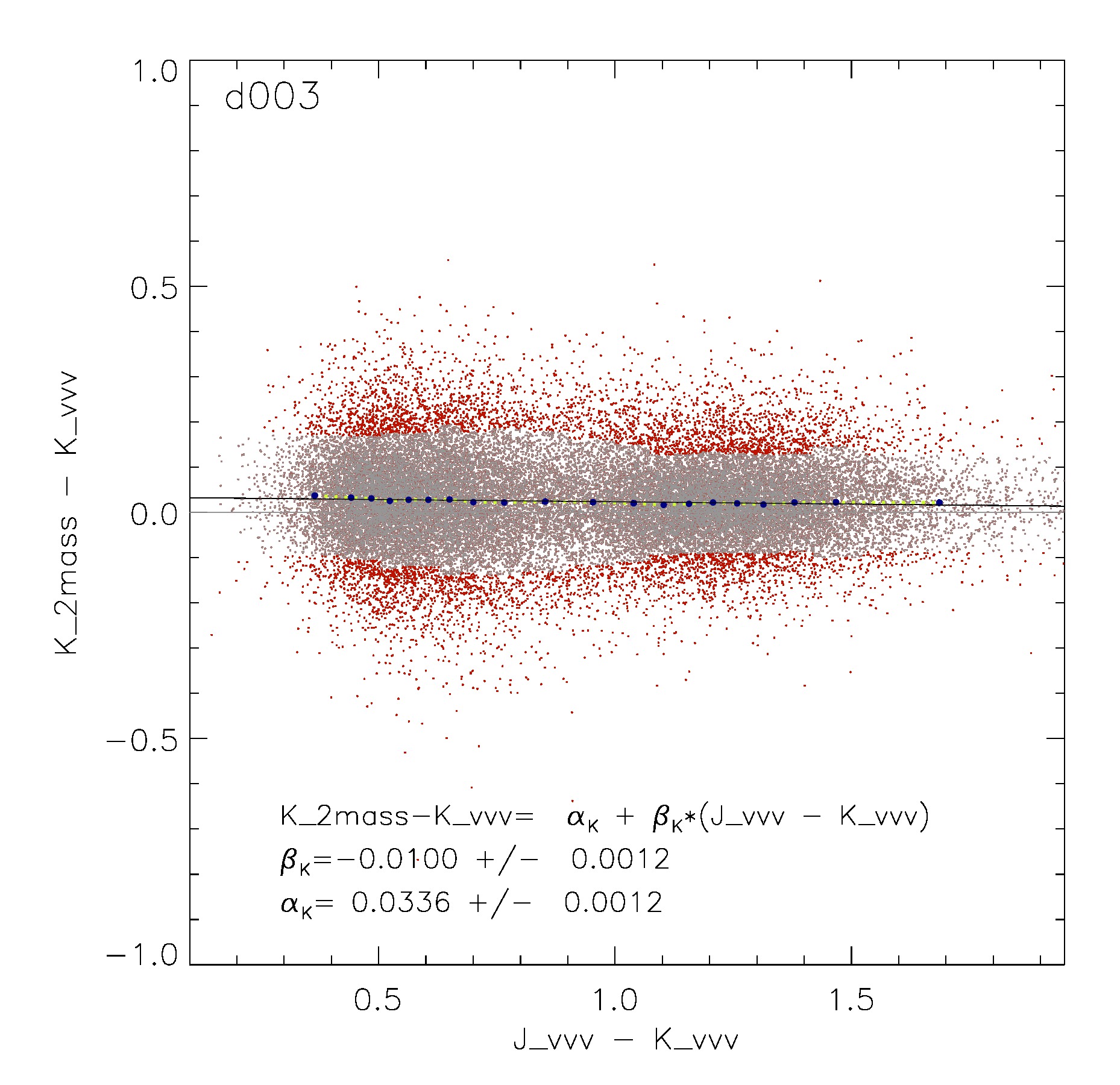}
  \includegraphics[width=8.0cm]{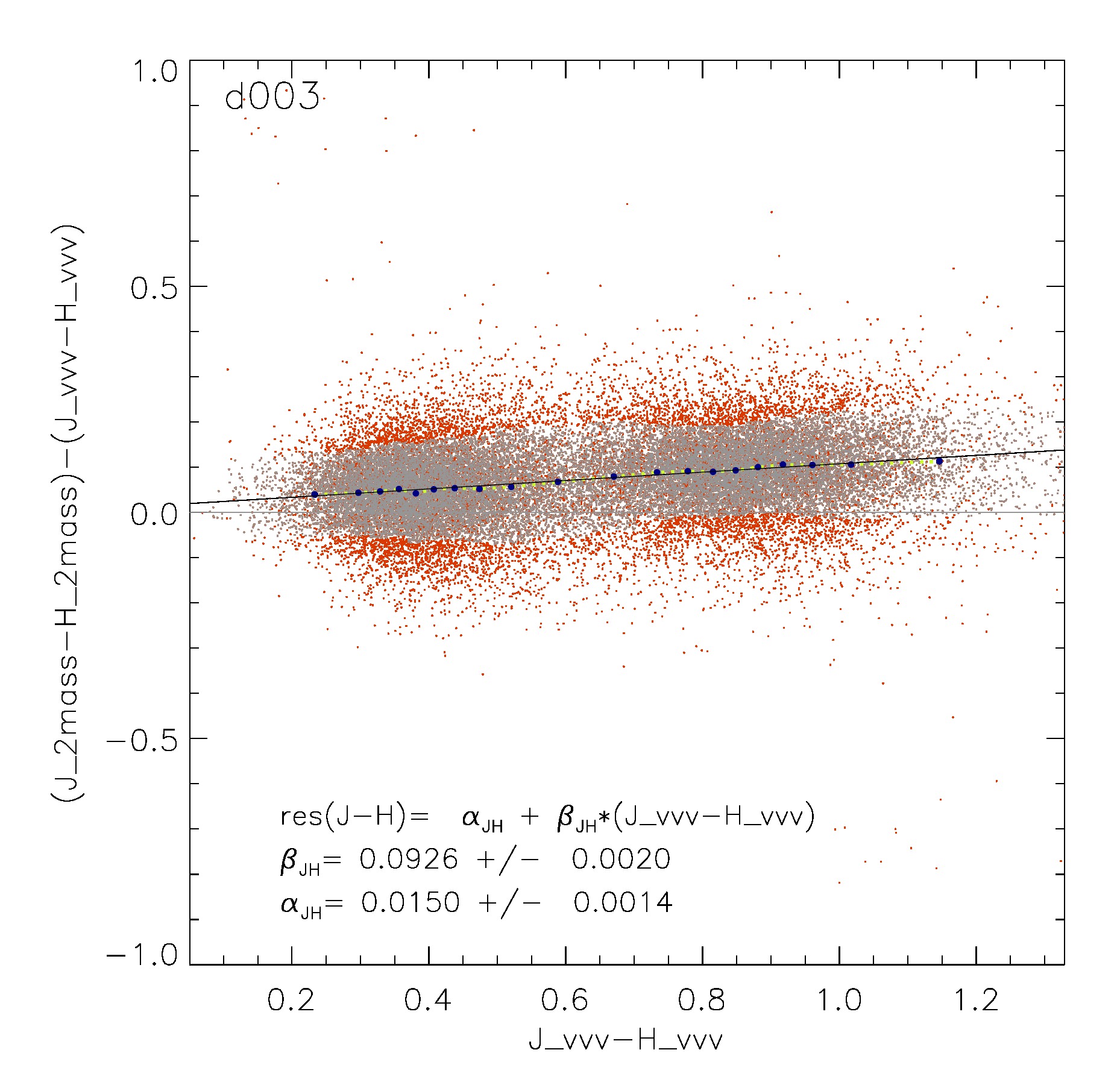}\\
  \includegraphics[width=8.0cm]{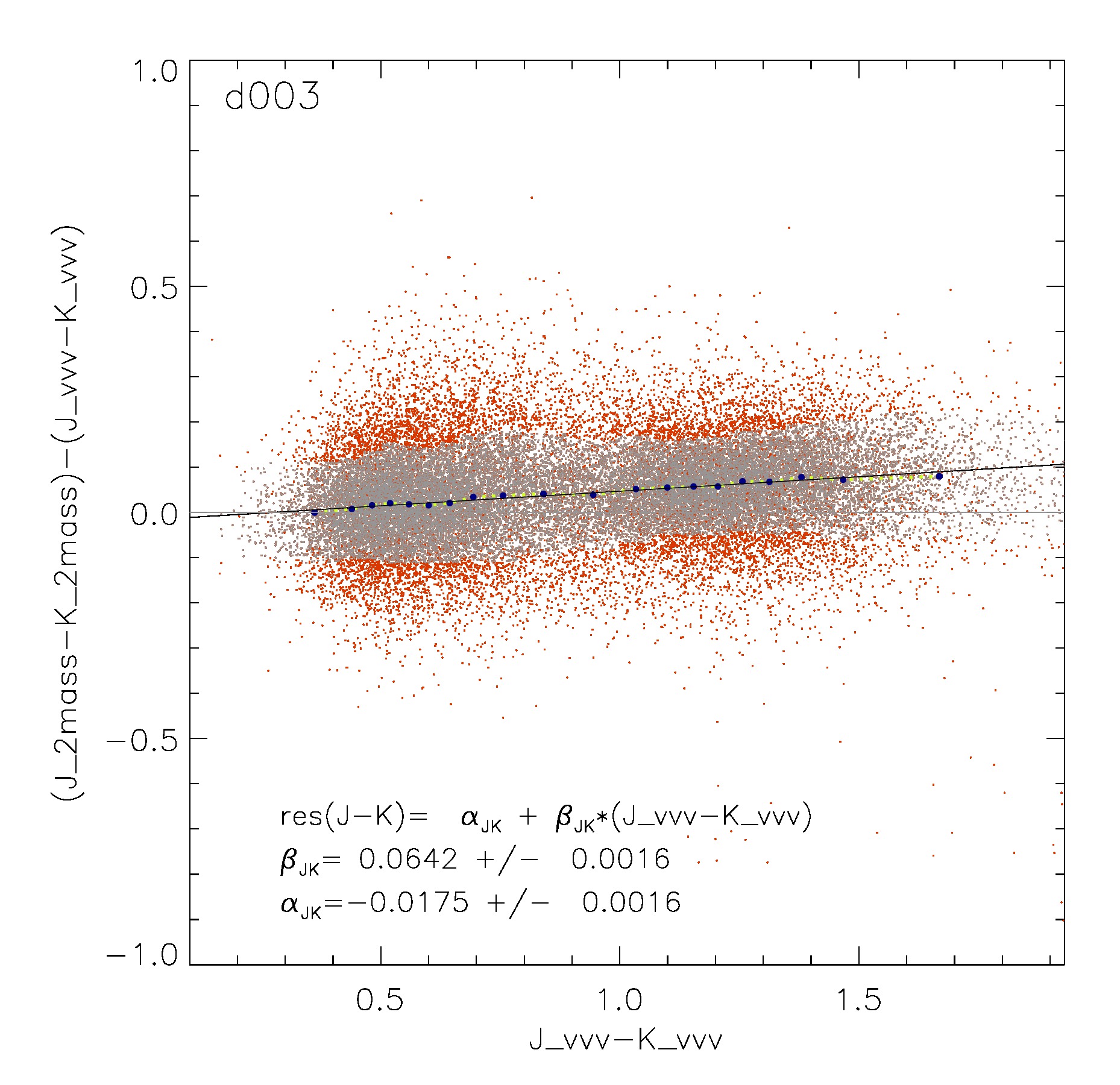}
  \includegraphics[width=8.0cm]{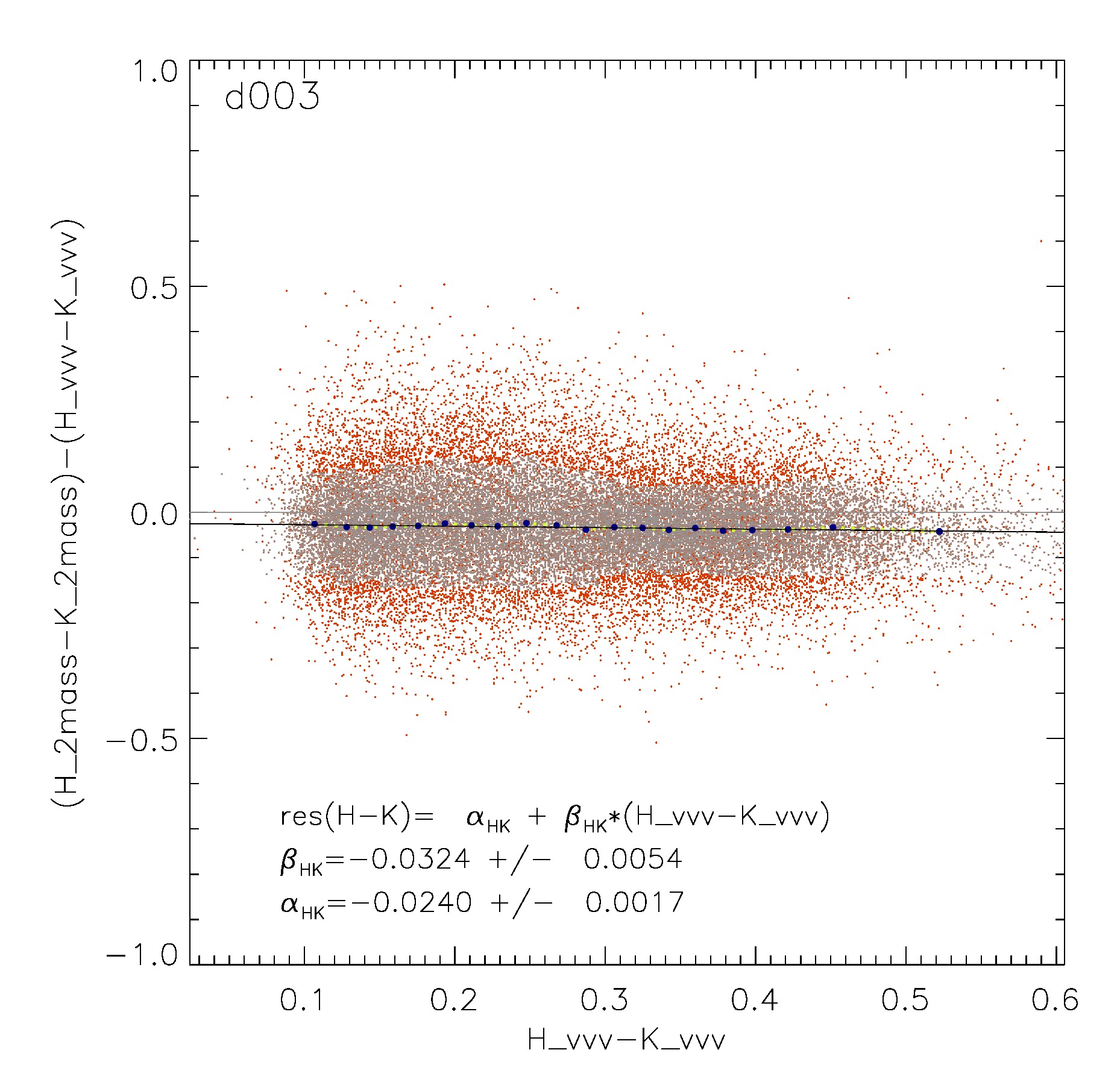}\\
   \caption{Comparison of 2MASS and VVV photometry for stars observed in tile/field 
     \emph{d003}. In each
     case an iterative clipping algorithm has been applied to each one
     of the 20 adaptive bins of the distribution. The selected stars
     of the clipping algorithm 
     (\emph{grey}) have been used to calculate the linear fit, and 
     the individual photometric uncertainties of each star were considered. }
              \label{fig:cmd1}%
    \end{figure*}

Zeropoints on the VISTA system are determined 
using the 2MASS data following a procedure similar to that described for 
WFCAM1 (Hodgkin et al. 2005). A color selected set of 2MASS stars
lying in each pawprint are chosen and their magnitudes on the VISTA photometric system are calculated
using the following color equations for data release 1.1 
(we adopt $K=K _{\rm s}$ in the equations for simplicity):

\begin{eqnarray}
J_{VVV} &=& J_{2MASS} - 0.077 (J_{2MASS} - H_{2MASS}), \nonumber \\
H_{VVV} &=& H_{2MASS} + 0.032 (J_{2MASS} - H_{2MASS}), \nonumber \\
K_{VVV} &=& K_{2MASS} + 0.010 (J_{2MASS} - K_{2MASS}),  \nonumber 
\end{eqnarray}

An extinction correction, based on  Schlegel, Finkbeiner \& Davis (1998; henceforth SFD), is applied according to the prescription
of Bonifacio, Monai \& Beers (2000).  The corrections can be found in  the CASU website\footnote{ http://casu.ast.cam.ac.uk/surveys-projects/vista/technical/photometric-properties}.  The analysis presented here is based on the derived colors and magnitudes established by the CASU pipeline.  A detailed account of the CASU pipeline can be found in Irwin et al. (2004), and will not be repeated here.


\section{Procedures}

\subsection{Catalogue construction and 2MASS matching}


The VISTA and 2MASS photometric systems do not exactly match, as
expected given the observations were carried out at different sites
with different telescopes, IR cameras, detectors and
filters. Figure \ref{fig:filters} shows a comparison of the
  transmission curves for both photometric systems.
As discussed earlier we wish to
 determine, on a tile by tile basis, the transformations between the
 VISTA and 2MASS systems for VVV data in the galactic disk. That is equivalent to changing the CASU calibration for each tile.  Nevertheless, the revised transformations should be more robust since red objects will be included in the calibration, whereas the CASU calibration relies principally on blue stars. 
 

The first step in obtaining the transformations was to select a set of VVV and 2MASS observations
exhibiting solid photometry.  A series of constraints were placed on the 2MASS and VVV
 photometry to account for undesirable effects arising from
 crowding and saturation. Extended sources were excluded.
 The procedure used to obtain a 2MASS-VVV
 catalogue for each tile can be summarized as follows: 
\begin{enumerate}[(a)]
\item  Only sources with VVV $K_{\rm s}$ photometry   
defined as "stellar" (sources with a Gaussian sigma parameter between 
$0.9$ and $2.2$) were analyzed.
 This parameter is derived from the three intensity
 weighted second moments.  $K_{\rm s}$ photometry was chosen since the data extend
 deeper than $J$ or $H$ for sources in the Galactic plane. Accounting for crowding effects in 
 $K_{\rm s}$ provides a corresponding solution for the shallower $J$ and $H$ data.   
\item Using the new list of $K_{\rm s}$ photometry, sources in close proximity to each other are subsequently culled, that is, stars exhibiting $r<2\farcs0$ (i.e. the 2MASS pixel size) and whose magnitudes display less than a 2 magnitude differential with respect to the brightest star.
 \item The resulting catalogue is then matched with 2MASS,
 where only stars with photometric quality flag \emph{"A"} or
 \emph{"B"} in a radius of  $0\farcs3$ are selected.
For $JHK_{\rm s}$ data the 2MASS photometric quality flags \emph{"A"} and \emph{"B"}
 correspond to $SNR>10$ and $SNR>7$, respectively.
\item The final list is   
 constructed by cross-referencing the VVV $K_{\rm s}$ and 2MASS $JHK_{\rm s}$
 list, received from the previous step, with the rest of the VVV $J$
 and $H$ data matched using a radius of $0\farcs1$. 
\end{enumerate} 

 Once a clean VVV-2MASS catalogue has been created for each VVV
 tile, the transformation equations were derived. The procedure
 is similar to that employed by Carpenter (2001).  Linear
fits for the variables were determined, namely:
 ($K_{\rm s}$)$_{2MASS}$-($K_{\rm s}$)$_{VVV}$ versus ($J-K_{\rm s}$)$_{VVV}$,
 ($J-H$)$_{2MASS}$ versus ($J-H$)$_{VVV}$,  
 ($J-K_{\rm s}$)$_{2MASS}$ versus ($J-K_{\rm s}$)$_{VVV}$, 
 and ($H-K_{\rm s}$)$_{2MASS}$ versus ($H-K_{\rm s}$)$_{VVV}$ and coefficients 
 ($\alpha_K,\beta_K$), ($\alpha_{JH},\beta_{JH}$), ($\alpha_{JK},\beta_{JK}$) and
 ($\alpha_{HK},\beta_{HK}$), respectively. 
 Thus, the derived linear fits correspond to the equations:

\begin{eqnarray}
&&K_{2MASS}-K_{VVV} = (J_{VVV} - K_{VVV})\ \beta_K + \alpha_K \ \ , \\
&&(J_{2MASS}-H_{2MASS})-(J_{VVV}-H_{VVV}) = (J_{VVV} - H_{VVV})\ \beta_{JH} \nonumber \\
&&\ \ \ \ \ \ \ \ \ \ \ \ \ \ \ \ \ \ \ \ \ \ \ \ \ \ \ \ \ \ \ \ \ \
\ \ \ \ \ \ \ \ \ \ \ \ \ \ \ \ \ \ \ \ \ \ \ \ \ \ \ \ + \alpha_{JH}\ \ ,\\ 
&&(J_{2MASS}-K_{2MASS})-(J_{VVV}-K_{VVV}) = (J_{VVV} - K_{VVV})\ \beta_{JK} \nonumber \\ 
&&\ \ \ \ \ \ \ \ \ \ \ \ \ \ \ \ \ \ \ \ \ \ \ \ \ \ \ \ \ \ \ \ \ \
\ \ \ \ \ \ \ \ \ \ \ \ \ \ \ \ \ \ \ \ \ \ \ \ \ \ \ \ + \alpha_{JK}\ \ ,\\ 
&&(H_{2MASS}-K_{2MASS})-(H_{VVV}-K_{VVV}) = (H_{VVV} - K_{VVV})\ \beta_{HK} \nonumber \\
&&\ \ \ \ \ \ \ \ \ \ \ \ \ \ \ \ \ \ \ \ \ \ \ \ \ \ \ \ \ \ \ \ \ \
\ \ \ \ \ \ \ \ \ \ \ \ \ \ \ \ \ \ \ \ \ \ \ \ \ \ \ \ + \alpha_{HK}\ \ .
\end{eqnarray}
\normalsize

 An iterative clipping algorithm was applied to reject stars beyond
 $2.5\sigma$ for each adaptive bin (i.e. uniformly populated bins).  That
  allowed us to establish a robust determination of the coefficients for
 the photometric transformation in each case.  A
 limiting magnitude was applied to each filter
 during the calculation of the transformations, which ensures that saturated photometry was avoided. 
 The limiting magnitudes used during the procedure were
 $13.8$, $12.8$ and $12.8$, for $J$, $H$ and $K_{\rm s}$ respectively.
 An example of the color-color and color magnitude diagrams (CMD) 
 in both photometric systems for tile \emph{d003} is shown in Fig. \ref{fig:cmd0}, whereas 
 the result of the fitting procedure is shown in Fig.  \ref{fig:cmd1}.

\begin{table}[b]
\begin{minipage}{\linewidth}
\renewcommand{\footnoterule}{}
\caption{Correlation coefficients between 2MASS-VVV transformation
  coefficients and E(B-V)}             
\label{table:rscoef}      
\centering                          
\begin{tabular}{l r l}        
Coeff &  $\ \ r_S\ $  \footnote{Spearman's correlation coefficient} \ \
\ & $Prob(r_S)$ \footnote{Significance of the correlation} \\    
\hline                        
$\alpha_K$   &     0.289   & 3.026e-04 \\
$\beta_K$    &    -0.395  & 4.425e-07 \\
$\alpha_{JH}$ &    0.424   &  5.209e-08  \\
$\beta_{JH}$  &   -0.549   & 2.562e-13  \\
$\alpha_{JK}$  &  -0.183   &  2.397e-02 \\
$\beta_{JK}$  &   -0.115  &   1.584e-01  \\
$\alpha_{HK}$ &   -0.692  & 5.682e-23  \\
$\beta_{HK}$ &    0.708   & 1.886e-24   \\
\hline                                   
\end{tabular}
\end{minipage}
\end{table}

\section{Discussion}
\begin{figure}
   \centering
  \includegraphics[width=8cm]{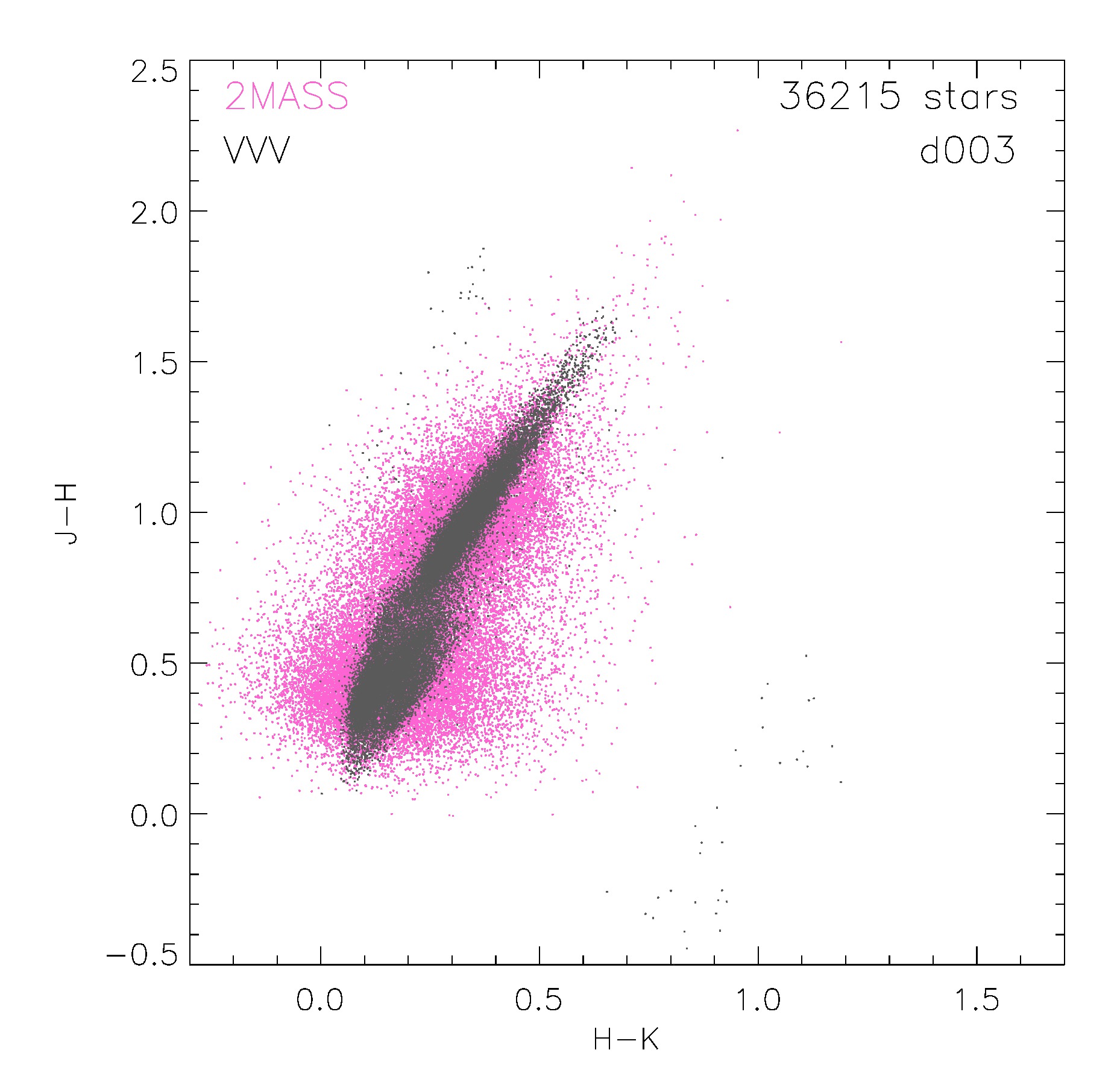}
   \caption{Color-color diagram for VVV stars transformed to the 2MASS
     photometric system for tile \emph{d003}. The transformation equations derived in 
     Fig. \ref{fig:cmd1} have been applied to the VVV colors (grey).  The same stars featured in the 2MASS catalogue are overplotted (pink). 
   }
              \label{fig:cmd2}%
    \end{figure}

 Figure \ref{fig:cmd2} displays the color-color diagram of tile \emph{d003}, where the
 calculated transformations were applied to the respective VVV colors. 
 The Gaussian fitting applied to the histograms of the  $JHK_{\rm s}$ magnitude residuals, in
 our transformations for tile \emph{d003},  exhibits $\sigma \simeq
 0.05\ mag$ for stars in the upper $25\%$ of the
   magnitude range used to
 calculate the photometric transformations. These residuals are
 dominated by the 2MASS magnitude dispersion, where 2MASS photometric
 errors are typically several times higher ($\sim 6$ on average for
 this tile) than
 those in the VVV catalogues.
 
 
 Table \ref{table:tablecoef1} lists
 the coefficients obtained for the 152 VVV tiles, while 
 Fig. \ref{fig:megatable1} displays the same coefficients as 
 a function of Galactic longitude and latitude.  
  Similarly, tables \ref{table:tablecoef1ms} and
  \ref{table:tablecoef1rgb} show the derived coefficients per tile for
  subsamples dominated by main sequence and post main sequence stars 
 respectively.
 At first glance, the figures suggest a non-random behavior that is presumably related to the
 structure of the Milky Way. In order to test that hypothesis, we compared how the coefficients varied for  low
(\emph{red}, $|b|\lesssim 1^{\circ}$) and high Galatic latitude
 fields (\emph{green}, $ 2.1^{\circ} \gtrsim |b| \gtrsim 1^{\circ}$).  For each
 subsample we fitted a fourth-order polynomial. 
%
%
 A clear distinction
 between high- and low-latitude
 fields is observed in  the photometric coefficients, with the
 apparent exception of the $\beta_{JK}$ parameter.
 A similar analysis can be drawn by dividing the sample in low longitude
   (\emph{red}, $l \geq 320^{\circ}$)
   and high longitude fields (\emph{green}, $  l <
   320^{\circ}$). Where we have fitted a third-order polynomial to
 fit the subsamples in each case.
 
The variations of the transformation coefficients across
 the Galaxy may be caused by multiple effects, as discussed below.

\subsection{Extinction on the disk for the VISTA fields}
 Extinction in the Galactic plane can be extreme and uneven at small
 scales. As mentioned, an extinction correction was employed in the VISTA pipeline based in part on
 the SFD map.  
 The problems of SFD in regions of high extinction are well
 documented. Arce \& Goodman (1999) evaluated the reliability of the
  SFD maps in the \object{Taurus Dark Cloud} complex, using 4 separate methods. 
Their results demonstrate a consistent overestimation by a factor of 1.3
 to 1.5 in regions of smooth extinction and $A_V > 0.5$ mag.  By contrast,
 reddening values were underestimated in regions with steep extinction gradients. Subsequent studies have shown similar results in 
 globular clusters (see Majewski et al. 2011 and references therein), whereby
 reddening values were overestimated by factors of 1.2 to 1.5. 
 Moreover, the comparison between Majewski et al. (2011)  extinction
 map, based on 2MASS (NIR)/Spitzer-IRAC
 observations, and the SFD map revealed clear discrepancies. 
 Majewski et al. (2011) attributed the offset to the fact that long wavelength
 (100 $\mu$)  infrared dust emission is not a viable tracer of dust extinction (SFD maps).  
 A similar result was found recently using VVV data. Gonzalez
 et al. (2012) compared their bulge extinction map with the 
 SFD map, and a significant difference appeared for $|b|<6^{\circ}$.
 In addition to the limitations of the applied SFD maps, offsets in the photometric
 transformations are expected owing to the different 
 filters employed by 2MASS and VVV.  Thus the transformations may be 
 affected by reddening and spectral type. Figure \ref{fig:isoch}
 illustrates that effect via a comparison of Padova isochrones
 (Girardi et al. 2000, 2002) on the 2MASS and VISTA photometric
 systems. In that example an old
 disk population ($\sim\ 10\ Gyr$; Carraro et al. 1999)
 affected by high extinction  ($A_V \sim 10$) displays divergent colors.

 \begin{figure*}
   \centering
  \includegraphics[width=17cm]{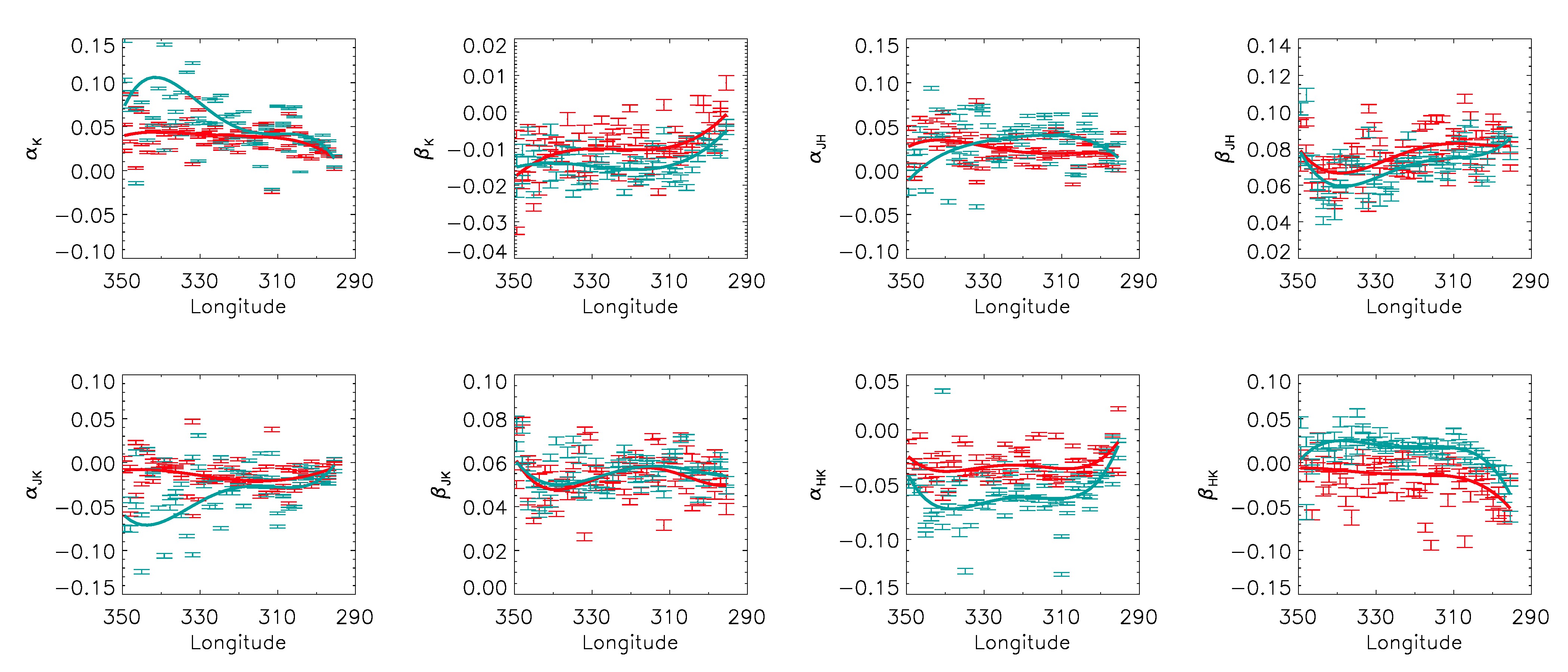}\\
 \includegraphics[width=17cm]{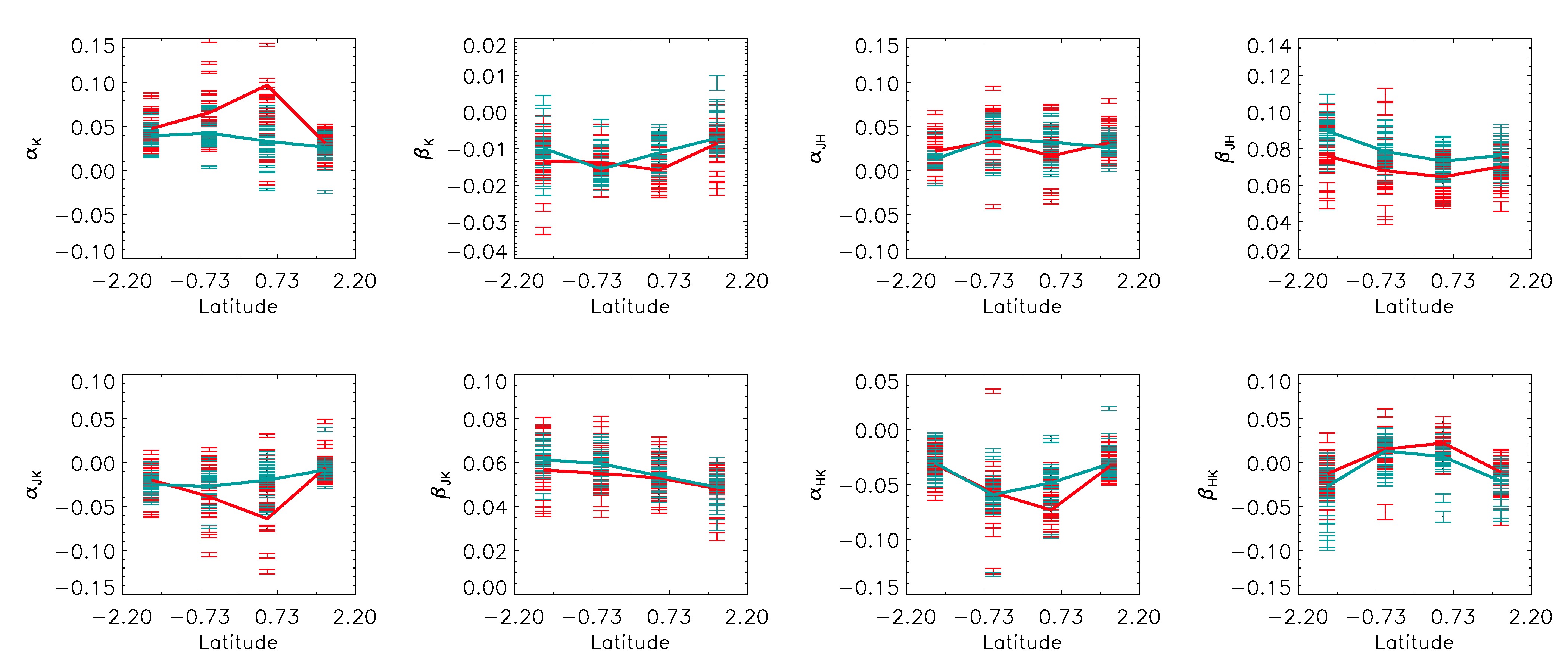}
   \caption{Coefficients of the VVV-2MASS transformations as a
     function of Galactic longitude and latitude for 152 VVV tiles in the
     Galactic disk. 
    \emph{Top and second row,} we divided the sample into low (\emph{red};
      $|b|\lesssim 1^{\circ}$) and high latitude
      fields (\emph{green}; $ 2.1^{\circ} \gtrsim |b| \gtrsim 1^{\circ}$). 
     A fourth order polynomial was fitted in each case. \emph{Third
       and bottom row}, photometric coefficients divided into low longitude
     (\emph{red}; $l \gtrsim 320^{\circ}$), and high longitude fields (\emph{green}; $l \lesssim 320^{\circ}$)      
 }
              \label{fig:megatable1}%
    \end{figure*}


 Figure \ref{fig:extcor}  and Table \ref{table:rscoef} show the
 photometric coefficients for 152 tiles as a function of the reddening used in the
 zeropoint correction for each tile, and their respective correlation
 results ($r_S$ is the Spearman's rank order correlation coefficient).  
 The calculated correlations are significant in most of  the
 coefficients, only  
 $\beta_{JK}$ and $\alpha_{JK}$ show little dependence on $E(B-V)$. Thus these results confirm our 
 original assessment regarding the influence of extinction (see also Fig. \ref{fig:megatable1}).  
 Since 2MASS
 and VVV colors and magnitudes must coincide for 
 $A0V$ spectral type and $E(B-V)=0$,  as Fig. \ref{fig:isoch} shows, our
 photometric transformations should follow consistent relations when
 extrapolated to zero reddening. 
 As expected,
  the linear fits for the $\alpha$ coefficients tend to zero for
  $E(B-V)=0$, an effect that seems to grow stronger with the correlation $r_S$. 
  Similarly, and in spite of the dispersion observed, these plots show a rough agreement with the inverse
  transformations that can be derived from the CASU equations (without their extinction corrections) in Section
  2.2.; which produce $\beta_{JK} = 0.075$ and  $\beta_{JH} = 0.109$
  for $E(B-V)=0$, as can be seen in Fig. \ref{fig:extcor}. All this 
  confirms the reliability of the photometric transformations obtained.

\subsection{Mapping the Galactic disk with VVV}

 Figure \ref{fig:cmd140} hosts the
 color-magnitude and color-color diagram for all fields in the VVV
 catalogue with VVV-2MASS color transformations.  The diagram features 88 million stars
 obtained from our combined JHK$_{\rm s}$ catalogues of the Galactic
 disk. 
 The combined CMD reveals the saturated population around 
 $K_{\rm s} \sim 10\ mag$.  However, our tests with individual
 tile-catalogues implied that saturation was typically near  
 $K_{\rm s}\sim 13\ mag$ (Fig. \ref{fig:cmd1}). 

The combined color-color diagram can be used to calculate
 the infrared color excess ratio (Indebetouw et al. 2005). The
 measured color excess ratio in 
 our diagram is $E(J-H)/E(H-K) = 2.13 \pm 0.04$, which was inferred from the VVV data converted
to the 2MASS system with $1.5\geq (H-K_{\rm s})\geq 0.5$.   The corresponding value in the original VISTA system is 
 $E(J-H)/E(H-K) = 2.02 \pm 0.04$.   These reddening laws are in general agreement with 
 previously reported values for numerous lines of sight toward the inner
Galaxy (Strai\v{z}ys and Laugalys 2008; Majaess et al. 2011).

  \begin{figure}
   \centering
  \includegraphics[width=8.0cm]{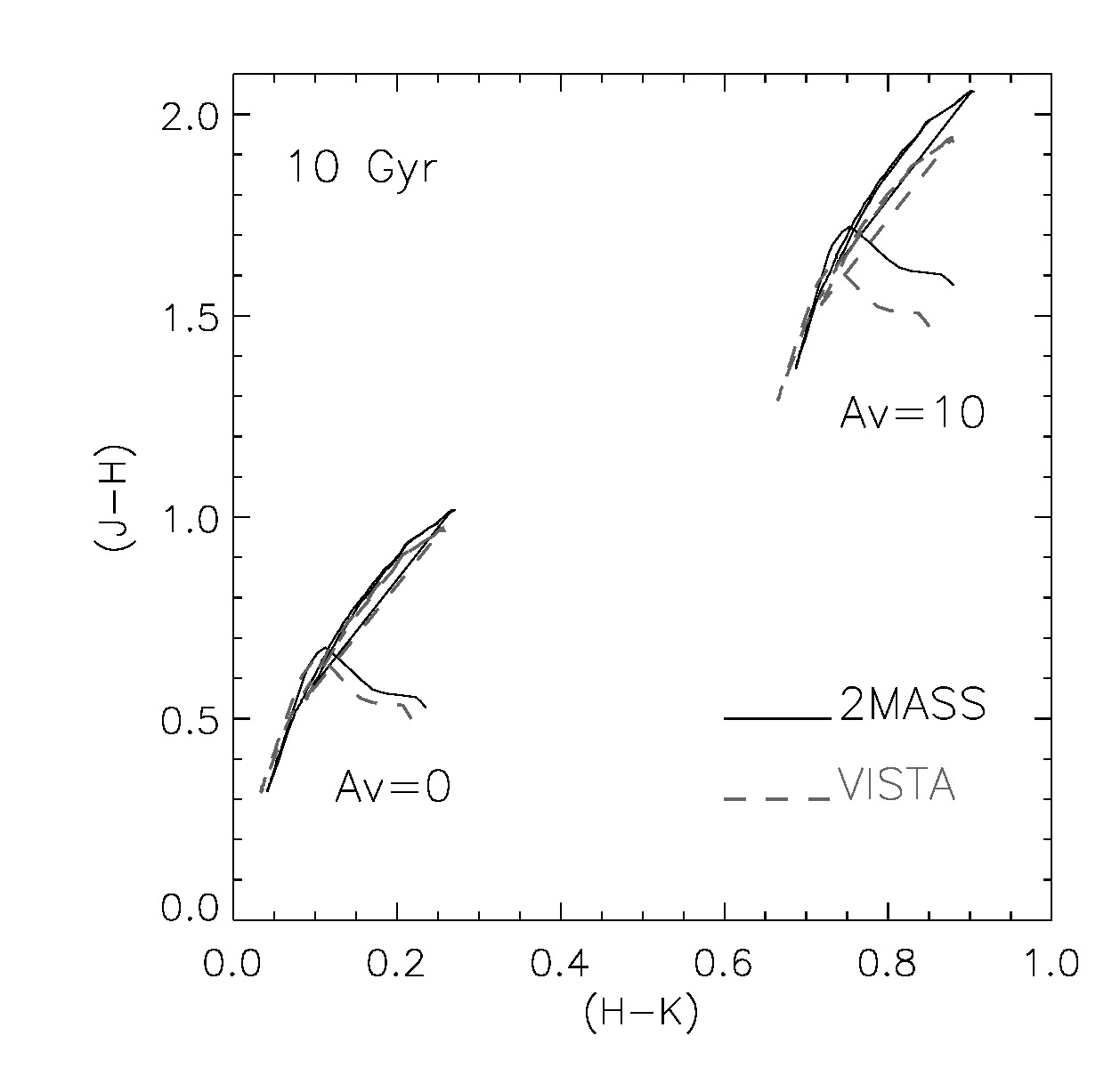}
   \caption{Comparison of isochrones in the 2MASS (black) and VISTA (grey)
     photometric systems for a 10 $Gyr$ population ($Z=0.019$) observed through two extinction values (Av=0 and
     Av=10).  
   }
              \label{fig:isoch}%
    \end{figure}

 Figure \ref{fig:lbmap41}  shows the source-count
 maps for all tiles processed for this work. 
 Duplications in the overlapping regions between
 tiles have been avoided in the starcount maps, as in the CMD and
 color-color diagrams, by constructing simultaneously the three
 corresponding binned plots (CMD, color-color diagram and
 starcount map). Once a pixel has been used in the starcount map, only
 counts from the same tile will be accepted in the three binned plots.
 The first of these count maps in Fig. \ref{fig:lbmap41} includes all $136\times10^6$ detected sources
 in the 152 tiles, regardless of their classification. 
 The second map consists only of stellar sources ($88\times10^6$ objects).
There exist differences in the general number of counts per tile 
 which we attribute to variations in the observation conditions
 between the tiles, in addition to patchy disk obscuration. Similarly, a marginal vertical stripe
 pattern is observed in many tiles.  That pattern is a known
 background variation related to the construction of the tiles from
 the 6 pawprint images. 
 As expected, when compared with the map including just stellar sources
 (\emph{second row}), the general map including all the sources (\emph{top})
displays more detailed structure in regions where diffuse sources are 
 expected.  Finally, for the last two starcount maps, we selected two
 subsamples of all the stellar sources by defining
 the following region in the color-color diagram:
\begin{eqnarray} 
 (H-K)\times 1.97 + 0.54 \geq (J-H) \geq (H-K)\times 1.97 + 0.24, \nonumber
\end{eqnarray}
 and dividing the stars in that strip at $(H-K)=0.4$.  Stars featuring
 $(H-K) \geq 0.4$ should be dominated by disk giants with moderate to
 high extinction, while stars exhibiting $(H-K)<0.4$ are dominated by
 nearby disk giants and dwarfs with low extinction. The resolution and
 extent of these maps allow for a detailed study of Galactic
 structure which will be the subject of a future work.
   


   \begin{figure*}
   \centering
  \includegraphics[width=15cm]{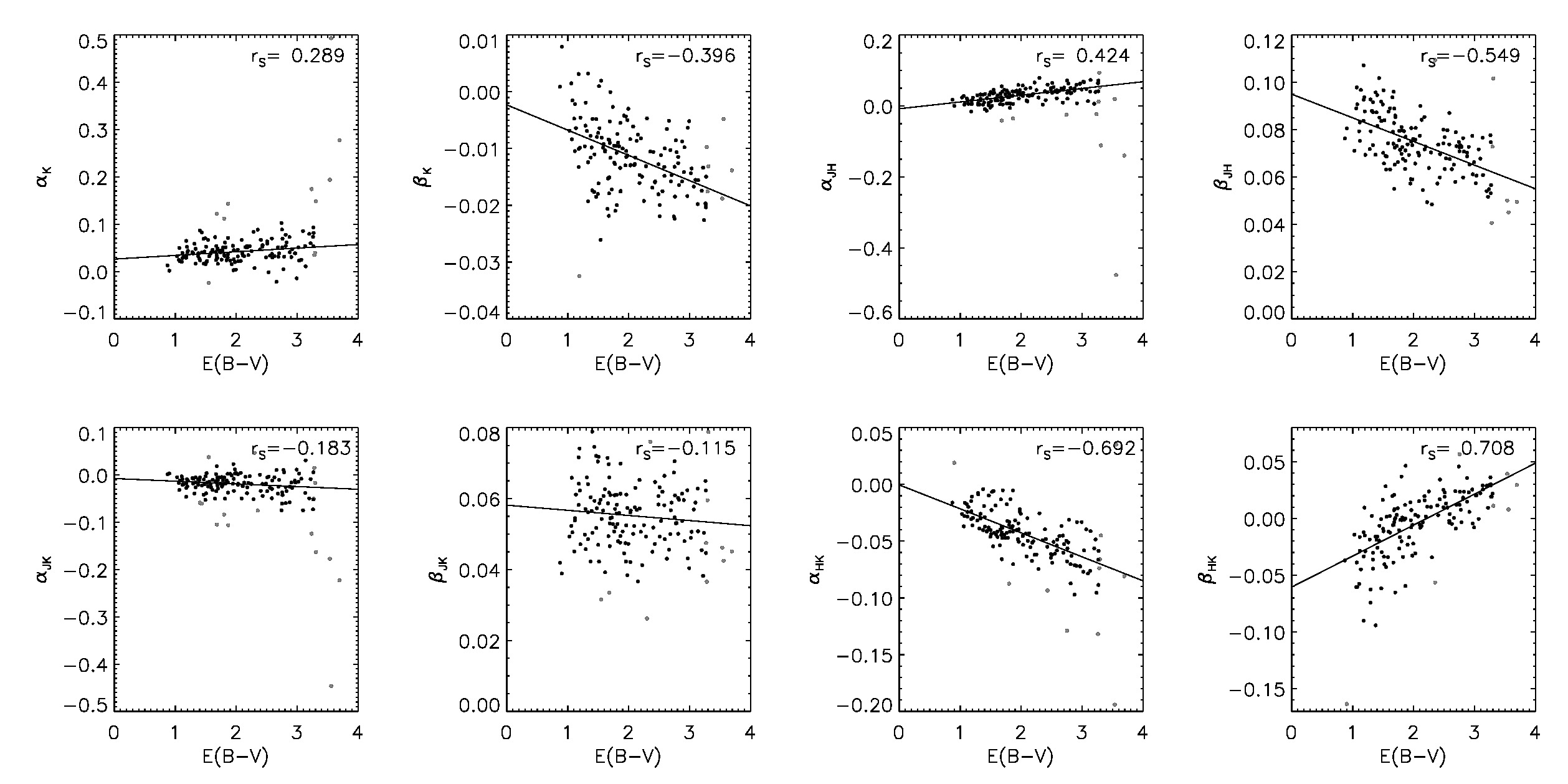}
   \caption{Coefficients of the VVV-2MASS transformations as a
     function of the reddening used to correct the zeropoint for 152 VVV fields in the
     Galactic disk. We included in each plot a linear fit with an
     iterative clipping algorithm similar to that used to calculate
     the photometric transformations. Grey points 
     are those rejected during the clipping procedure.
     Moreover, the Spearman's rank order correlation coefficient $r_S$ has
     been calculated in each case. }
              \label{fig:extcor}%
    \end{figure*}

  \begin{figure*}
   \centering
  \includegraphics[height=12cm]{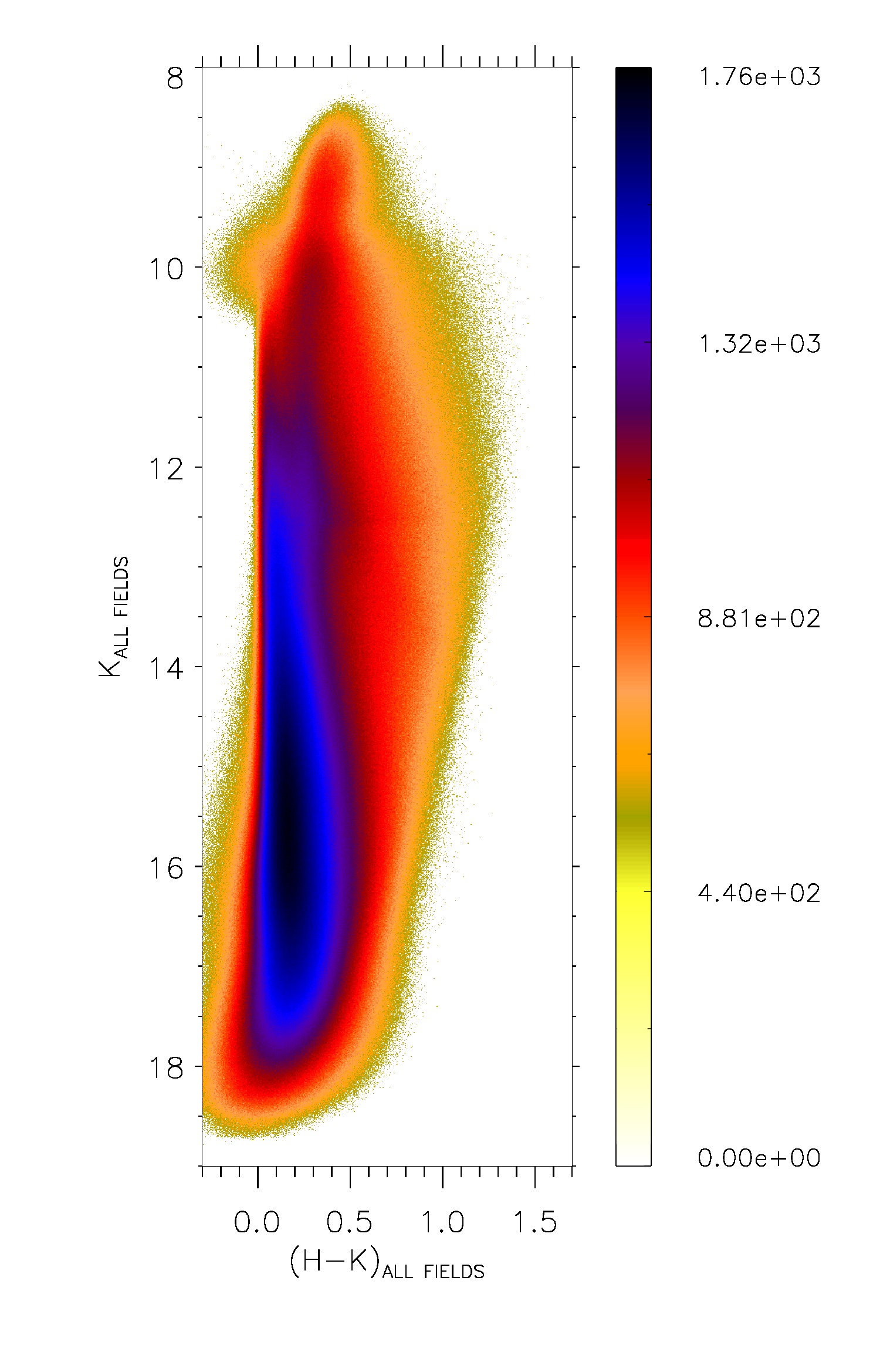}
  \includegraphics[height=12cm]{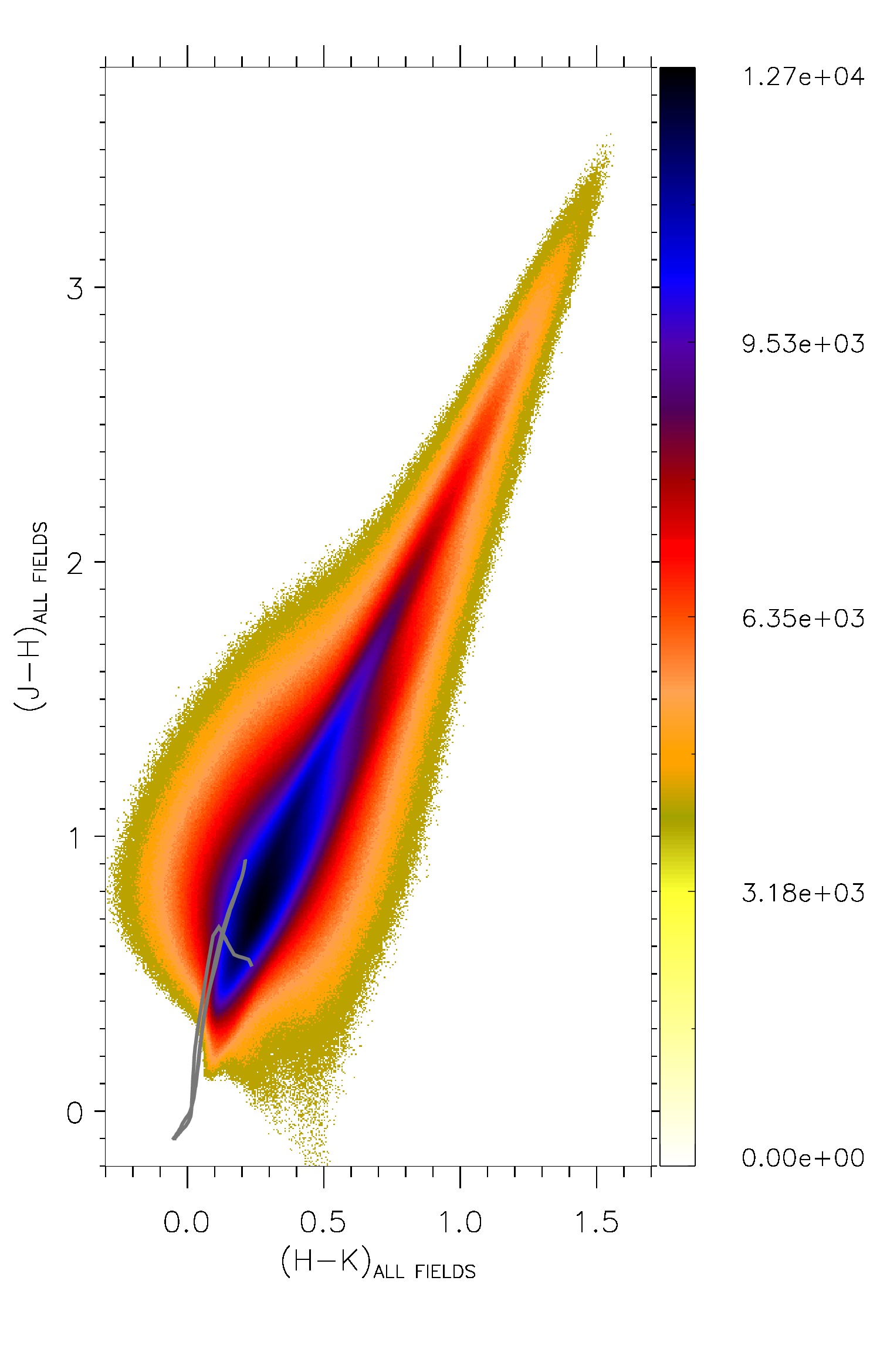}
   
   \caption{Binned color-magnitude and color-color diagrams for all sources defined as
     stellar in the VVV catalogues (86 millions), which constitutes 152 
     tiles of the Galactic disk. Magnitudes have been
     transformed to the 2MASS photometric system in each tile using the respective coefficients. 
     The color-magnitude diagram has been calculated at a
     resolution of 500$\times$1600 bins, which corresponds to a
     binsize of 0.005 mags/bin. Similarly, our color-color diagram 
     for the tiles of the Galactic disk was constructed at a
     resolution of 400$\times$800 bins and the same binsize of the
     binned CMD. The latter includes an unreddened 0.05 Gyr isochrone (grey solid
     line) in the
     2MASS system for reference.}
              \label{fig:cmd140}%
    \end{figure*}

\begin{landscape}
  \begin{figure}
   \centering
   \includegraphics[width=24cm]{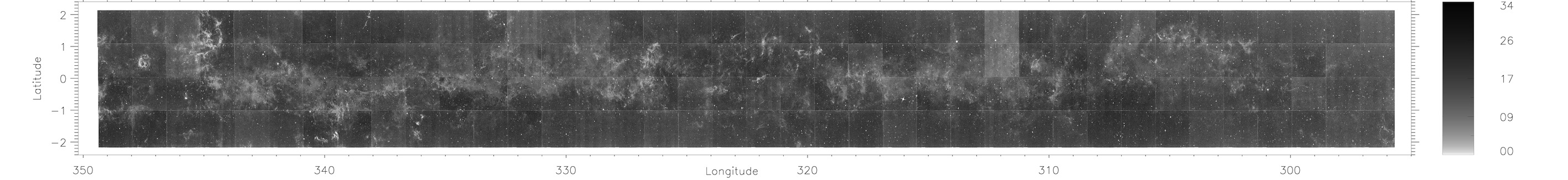}
   \vspace{1cm}\\
   \includegraphics[width=24cm]{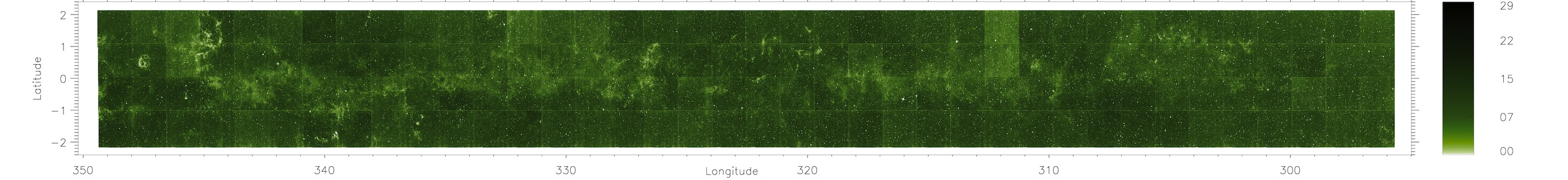}
   \vspace{1cm}\\
   \includegraphics[width=24cm]{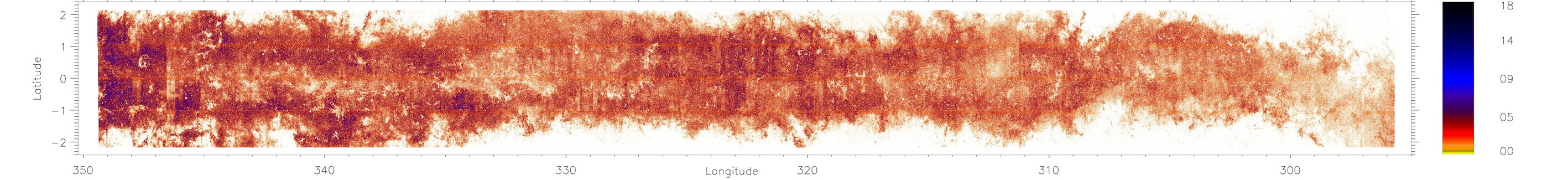}
   \vspace{1cm}\\
   \includegraphics[width=24cm]{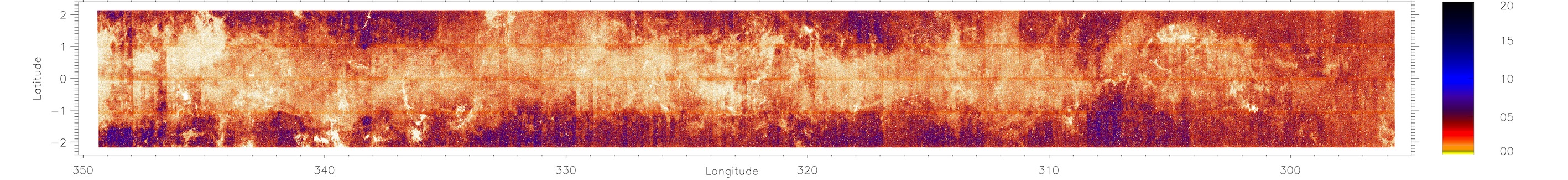}
   \caption{Map featuring the number of sources or the disk tiles (152 fields) in
     VVV with a 0.005$^{\circ} \times$0.005$^{\circ}$ bin/pixel
     size. \emph{Top},  map for all the sources in our combined
     JHK$_{\rm s}$ catalogues, 136
     million sources. \emph{Second row}, map for all the stellar
     sources, 88 million stars.
     \emph{Third row}, map for stellar sources for the
      objects lying  
      in a stripe in the color-color diagram defined by $ (H-K)\times 1.97 + 0.54 \geq (J-H)
      \geq (H-K)\times 1.97 + 0.24  $ and $(H-K) \geq 0.4$; this
      selection should be dominated by disk red giants with high
      extinction. \emph{Bottom,} Same as before, but for $(H-K) \leq 0.4$;   
      the population selected in this starcount map should be
      dominated by red-giants and some main sequence stars with low 
      extinction.
}
              \label{fig:lbmap41}%
    \end{figure}
\end{landscape}

\section{Conclusions}


We have derived empirical transformations from VVV to
2MASS for 152 fields of the VVV survey of the Galactic disk.
The transformations in each case have been derived using
an iterative clipping algorithm, which improves the robustness
of the coefficients. The coefficients reflect 
the inverse of the relations used in calibrating onto the VISTA photometric system, and
as expected we have found statistically significant
correlations between the transformation coefficients and the
Galactic extinction used in the disk. Our results also suggest some scatter
in the transformations which in the case of high extinction fields
 seem to be related with the inadequacy of the SFD maps used
 in the zeropoint calibration and require further analysis.
 Our photometric transformations allow to avoid some of the described
 uncertainties when working with the VVV catalogues as well as to
 complement with 2MASS observations when working with saturated objects in the VVV catalogues.
  
 In addition, we presented a stellar CMD and a CCD for $134\times10^6$ sources in the Galactic plane. The stellar CMD is dominated by main sequence stars in the disk, whose breadth is widened by differential extinction. The sequence tied to more distant red giants is also seen.  In addition, the derived infrared color excess ratio is in agreement with previously reported values. Finally, we present density maps of main sequence stars and red giants. These are useful for identifying overdensities such as star clusters and Galactic spiral arms, as well as the less-populated regions that may correspond to dense clouds.


%


\begin{acknowledgements}
MS acknowledges support by Fondecyt project No. 3110188 and Comit\'e
Mixto ESO- Chile. 
We gratefully acknowledge use of data from the ESO Public Survey programme ID 179.B-2002 taken with the VISTA telescope, data
products from the Cambridge Astronomical Survey Unit, and funding from
the FONDAP Center for Astrophysics 15010003, the BASAL CATA Center for
Astrophysics and Associated Technologies PFB-06, the MILENIO Milky Way
Millennium Nucleus from the Ministry of Economy’s ICM grant
P07-021-F. RKS and DM acknowledge financial support from CONICYT
through Gemini Project No. 32080016 and by Proyecto FONDECYT Regular
No. 1090213. Support for RKS is provided by the Ministry for the
Economy, Development, and Tourism's Programa Iniciativa Cient\'{i}fica
Milenio through grant P07-021-F, awarded to The Milky Way Millennium
Nucleus. JB is supported by FONDECYT  No.1120601.
RK acknowledges support from Proyecto FONDECYT Regular No. 1130140 and Centro de Astrof\'{\i}sica de Valpara\'{\i}so.
ARL thanks partial financial support from DIULS project CDI12141. 
RB thanks financial support from FONDECYT Regular No. 1120668.
This material is based upon work supported in part by the
National Science Foundation under Grant No. 1066293 and the hospitality of the Aspen Center for Physics.      
\end{acknowledgements}

\begin{appendix}
\section{Coefficients of the Photometric Transformations}
\onecolumn
\tiny
\begin{landscape}
\centering
\begin{longtable}{ l c c l c c c c c c c c } 
\caption{Coefficients for the Photometric Transformations of 152 Disk tiles} \label{table:tablecoef1}\\        
\hline  \hline      
Tile & $RA$ & $DEC$ & $N_{fit}$& $\alpha_K$ & $\beta_K$ & $\alpha_{JH}$ & $\beta_{JH}$ & $\alpha_{JK}$ & $\beta_{JK}$ & $\alpha_{HK}$ & $\beta_{HK}$ \\ 
      &   (deg)  & (deg)   &       & (mag) &        & (mag)  &      & (mag) &      & (mag)    &    \\
\hline       
\endfirsthead 
\caption{Continued.} \\
\hline
Tile & $RA$ & $DEC$ & $N_{fit}$& $\alpha_K$ & $\beta_K$ & $\alpha_{JH}$ & $\beta_{JH}$ & $\alpha_{JK}$ & $\beta_{JK}$ & $\alpha_{HK}$ & $\beta_{HK}$ \\ 
      &  (deg)   & (deg)   &         & (mag) &       & (mag)  &       & (mag) &      & (mag)    &   \\    
\hline    
\endhead
\hline
\endfoot
\hline
\endlastfoot       
d001 & 175.8521 & -63.5298 & 33644 &  0.0157 $\pm$ 0.0010 & -0.0042 $\pm$ 0.0008 &  0.0430 $\pm$ 0.0013 & 0.0693 $\pm$ 0.0017 &  -0.0010 $\pm$ 0.0015 & 0.0446 $\pm$ 0.0013  &  -0.0402 $\pm$ 0.0015 & -0.0166 $\pm$ 0.0038 \\ 
d002 & 179.0511 & -63.8750 & 32106 &  0.0165 $\pm$ 0.0011 & -0.0001 $\pm$ 0.0011 &  0.0042 $\pm$ 0.0015 & 0.0860 $\pm$ 0.0021 &  -0.0108 $\pm$ 0.0016 & 0.0545 $\pm$ 0.0016  &  -0.0052 $\pm$ 0.0015 & -0.0652 $\pm$ 0.0048 \\ 
d003 & 182.3203 & -64.1487 & 36215 &  0.0336 $\pm$ 0.0012 & -0.0101 $\pm$ 0.0012 &  0.0150 $\pm$ 0.0014 & 0.0926 $\pm$ 0.0020 &  -0.0174 $\pm$ 0.0016 & 0.0642 $\pm$ 0.0016  &  -0.0237 $\pm$ 0.0016 & -0.0334 $\pm$ 0.0054 \\ 
d004 & 185.6456 & -64.3492 & 38982 &  0.0307 $\pm$ 0.0011 & -0.0064 $\pm$ 0.0011 &  0.0060 $\pm$ 0.0014 & 0.0968 $\pm$ 0.0020 &  -0.0200 $\pm$ 0.0016 & 0.0611 $\pm$ 0.0016  &  -0.0198 $\pm$ 0.0015 & -0.0449 $\pm$ 0.0051 \\ 
d005 & 189.0110 & -64.4743 & 39338 &  0.0291 $\pm$ 0.0014 & 0.0031 $\pm$ 0.0013 &  0.0057 $\pm$ 0.0017 & 0.0838 $\pm$ 0.0024 &  -0.0273 $\pm$ 0.0019 & 0.0582 $\pm$ 0.0019  &  -0.0305 $\pm$ 0.0020 & -0.0122 $\pm$ 0.0064 \\ 
d006 & 192.3968 & -64.5231 & 36247 &  0.0166 $\pm$ 0.0014 & 0.0032 $\pm$ 0.0013 &  0.0205 $\pm$ 0.0018 & 0.0873 $\pm$ 0.0024 &  -0.0159 $\pm$ 0.0019 & 0.0562 $\pm$ 0.0018  &  -0.0296 $\pm$ 0.0020 & -0.0303 $\pm$ 0.0063 \\ 
d007 & 195.7848 & -64.4954 & 41994 &  0.0411 $\pm$ 0.0012 & -0.0099 $\pm$ 0.0011 &  0.0171 $\pm$ 0.0015 & 0.0877 $\pm$ 0.0021 &  -0.0242 $\pm$ 0.0017 & 0.0623 $\pm$ 0.0016  &  -0.0361 $\pm$ 0.0017 & -0.0151 $\pm$ 0.0053 \\ 
d008 & 199.1573 & -64.3908 & 41225 &  0.0499 $\pm$ 0.0011 & -0.0084 $\pm$ 0.0012 &  0.0039 $\pm$ 0.0015 & 0.0978 $\pm$ 0.0022 &  -0.0320 $\pm$ 0.0016 & 0.0659 $\pm$ 0.0017  &  -0.0279 $\pm$ 0.0016 & -0.0388 $\pm$ 0.0059 \\
\end{longtable}
\end{landscape}
\normalsize

\section{Coefficients of the Photometric Transformations separated by
  Populations}
\begin{figure*}[h]
   \centering
  \includegraphics[width=15cm]{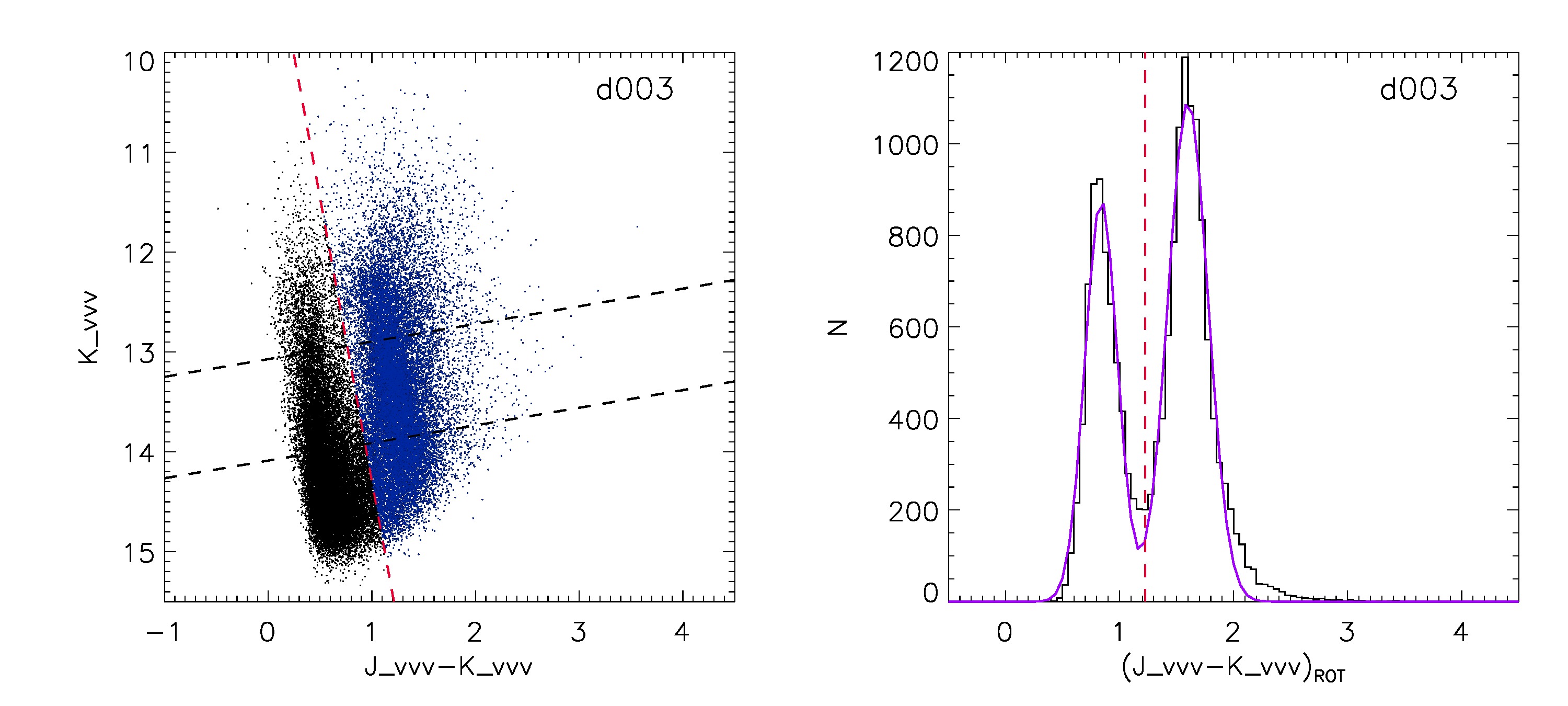}
   \caption{Example of the procedure applied to separate the stellar populations
     in each tile. \emph{Left}, 
     color magnitude diagram for the VVV tile \emph{d003}, the subsample
     of stars  between the black dashed lines has been chosen to
     calculate an histogram of the color equation $(J-K)_{ROT}=
     (J-K+1.0) \times 0.985 - (10.0 - K)\times0.174 $.   \emph{Right},
     respective histogram for the stars selected in the CMD, a double
     Gaussian fitting (solid purple line) is used estimate the minimum
     between the main sequence and post main sequence distributions (red
     dashed line), which is used to separate the populations. 
 }
              \label{fig:sepbypop}%
    \end{figure*}

 In addition to the transformations presented in the previous
  section, we have produced similar coefficients discriminating the 
  different stellar populations with a simple procedure. 
 Figure \ref{fig:sepbypop} illustrates
the technique: for each field we calculated a color histogram for a
range of colors, where a double Gaussian fit is used to estimate
the minimum between  the main sequence  and post-main sequence
star distributions.  This minimum is then used as limit to separate
both populations in the CMD.
Tables  \ref{table:tablecoef1ms} and
\ref{table:tablecoef1rgb} show the derived coefficients of the  photometric transformations
 for main sequence, and post main sequence stars respectively, which
 have been calculated using the same procedure applied to the complete
 sample.

\onecolumn
\tiny
\begin{landscape}
\centering
\begin{longtable}{ l c c l c c c c c c c c } 
\caption{Coefficients for the Photometric Transformations of 152 Disk tiles divided by population: main sequence stars} \label{table:tablecoef1ms}\\        
\hline  \hline      
Tile & $RA$ & $DEC$ & $N_{fit}$& $\alpha_K$ & $\beta_K$ & $\alpha_{JH}$ & $\beta_{JH}$ & $\alpha_{JK}$ & $\beta_{JK}$ & $\alpha_{HK}$ & $\beta_{HK}$ \\ 
      &   (deg)  & (deg)   &       & (mag) &        & (mag)  &      & (mag) &      & (mag)    &    \\
\hline       
\endfirsthead 
\caption{Continued.} \\
\hline
Tile & $RA$ & $DEC$ & $N_{fit}$& $\alpha_K$ & $\beta_K$ & $\alpha_{JH}$ & $\beta_{JH}$ & $\alpha_{JK}$ & $\beta_{JK}$ & $\alpha_{HK}$ & $\beta_{HK}$ \\ 
      &  (deg)   & (deg)   &         & (mag) &       & (mag)  &       & (mag) &      & (mag)    &   \\    
\hline    
\endhead
\hline
\endfoot
\hline
\endlastfoot       
d001 & 175.8521 & -63.5298 & 20830 &  0.0244 $\pm$ 0.0024 & -0.0180 $\pm$ 0.0033 &  0.0438 $\pm$ 0.0025 & 0.0653 $\pm$ 0.0053 &  -0.0108 $\pm$ 0.0030 & 0.0597 $\pm$ 0.0042  &  -0.0454 $\pm$ 0.0032 & 0.0103 $\pm$ 0.0127 \\ 
d002 & 179.0511 & -63.8750 & 18636 &  0.0266 $\pm$ 0.0027 & -0.0137 $\pm$ 0.0042 &  0.0149 $\pm$ 0.0028 & 0.0512 $\pm$ 0.0062 &  -0.0231 $\pm$ 0.0034 & 0.0756 $\pm$ 0.0053  &  -0.0152 $\pm$ 0.0031 & 0.0001 $\pm$ 0.0160 \\ 
d003 & 182.3203 & -64.1487 & 19237 &  0.0394 $\pm$ 0.0029 & -0.0178 $\pm$ 0.0045 &  0.0251 $\pm$ 0.0029 & 0.0619 $\pm$ 0.0067 &  -0.0280 $\pm$ 0.0036 & 0.0832 $\pm$ 0.0056  &  -0.0323 $\pm$ 0.0035 & 0.0208 $\pm$ 0.0169 \\ 
d004 & 185.6456 & -64.3492 & 18036 &  0.0238 $\pm$ 0.0032 & 0.0070 $\pm$ 0.0051 &  0.0222 $\pm$ 0.0032 & 0.0505 $\pm$ 0.0074 &  -0.0217 $\pm$ 0.0040 & 0.0649 $\pm$ 0.0065  &  -0.0289 $\pm$ 0.0037 & 0.0204 $\pm$ 0.0194 \\ 
d005 & 189.0110 & -64.4743 & 18075 &  0.0349 $\pm$ 0.0036 & 0.0001 $\pm$ 0.0051 &  0.0126 $\pm$ 0.0037 & 0.0602 $\pm$ 0.0075 &  -0.0396 $\pm$ 0.0046 & 0.0747 $\pm$ 0.0065  &  -0.0336 $\pm$ 0.0048 & 0.0220 $\pm$ 0.0214 \\ 
d006 & 192.3968 & -64.5231 & 16958 &  0.0088 $\pm$ 0.0035 & 0.0208 $\pm$ 0.0049 &  0.0271 $\pm$ 0.0037 & 0.0698 $\pm$ 0.0077 &  -0.0290 $\pm$ 0.0044 & 0.0771 $\pm$ 0.0063  &  -0.0372 $\pm$ 0.0046 & 0.0159 $\pm$ 0.0196 \\ 
d007 & 195.7848 & -64.4954 & 19198 &  0.0451 $\pm$ 0.0032 & -0.0130 $\pm$ 0.0046 &  0.0229 $\pm$ 0.0033 & 0.0675 $\pm$ 0.0070 &  -0.0395 $\pm$ 0.0041 & 0.0881 $\pm$ 0.0060  &  -0.0417 $\pm$ 0.0041 & 0.0319 $\pm$ 0.0187 \\ 
d008 & 199.1573 & -64.3908 & 20041 &  0.0403 $\pm$ 0.0029 & 0.0152 $\pm$ 0.0046 &  0.0128 $\pm$ 0.0031 & 0.0703 $\pm$ 0.0069 &  -0.0385 $\pm$ 0.0038 & 0.0756 $\pm$ 0.0059  &  -0.0302 $\pm$ 0.0039 & -0.0205 $\pm$ 0.0201 \\
\end{longtable}
\end{landscape}
\normalsize

\onecolumn
\tiny
\begin{landscape}
\centering
\begin{longtable}{ l c c l c c c c c c c c } 
\caption{Coefficients for the Photometric Transformations of 152 Disk tiles divided by population: post main sequence stars} \label{table:tablecoef1rgb}\\        
\hline  \hline      
Tile & $RA$ & $DEC$ & $N_{fit}$& $\alpha_K$ & $\beta_K$ & $\alpha_{JH}$ & $\beta_{JH}$ & $\alpha_{JK}$ & $\beta_{JK}$ & $\alpha_{HK}$ & $\beta_{HK}$ \\ 
      &   (deg)  & (deg)   &       & (mag) &        & (mag)  &      & (mag) &      & (mag)    &    \\
\hline       
\endfirsthead 
\caption{Continued.} \\
\hline
Tile & $RA$ & $DEC$ & $N_{fit}$& $\alpha_K$ & $\beta_K$ & $\alpha_{JH}$ & $\beta_{JH}$ & $\alpha_{JK}$ & $\beta_{JK}$ & $\alpha_{HK}$ & $\beta_{HK}$ \\ 
      &  (deg)   & (deg)   &         & (mag) &       & (mag)  &       & (mag) &      & (mag)    &   \\    
\hline    
\endhead
\hline
\endfoot
\hline
\endlastfoot       
d001 & 175.8521 & -63.5298 & 12814 &  0.0147 $\pm$ 0.0032 & -0.0032 $\pm$ 0.0019 &  0.0623 $\pm$ 0.0058 & 0.0524 $\pm$ 0.0052 &  0.0172 $\pm$ 0.0057 & 0.0331 $\pm$ 0.0034  &  -0.0483 $\pm$ 0.0045 & -0.0026 $\pm$ 0.0082 \\ 
d002 & 179.0511 & -63.8750 & 13469 &  -0.0129 $\pm$ 0.0033 & 0.0202 $\pm$ 0.0024 &  0.1162 $\pm$ 0.0053 & -0.0265 $\pm$ 0.0055 &  0.0211 $\pm$ 0.0056 & 0.0309 $\pm$ 0.0041  &  -0.0050 $\pm$ 0.0043 & -0.0669 $\pm$ 0.0102 \\ 
d003 & 182.3203 & -64.1487 & 16978 &  0.0118 $\pm$ 0.0034 & 0.0063 $\pm$ 0.0026 &  0.0456 $\pm$ 0.0054 & 0.0595 $\pm$ 0.0060 &  -0.0047 $\pm$ 0.0053 & 0.0542 $\pm$ 0.0041  &  -0.0271 $\pm$ 0.0042 & -0.0262 $\pm$ 0.0109 \\ 
d004 & 185.6456 & -64.3492 & 20946 &  0.0286 $\pm$ 0.0025 & -0.0050 $\pm$ 0.0020 &  0.0131 $\pm$ 0.0040 & 0.0901 $\pm$ 0.0045 &  -0.0229 $\pm$ 0.0039 & 0.0633 $\pm$ 0.0031  &  -0.0306 $\pm$ 0.0031 & -0.0195 $\pm$ 0.0083 \\ 
d005 & 189.0110 & -64.4743 & 21262 &  -0.0054 $\pm$ 0.0029 & 0.0291 $\pm$ 0.0023 &  0.0574 $\pm$ 0.0046 & 0.0264 $\pm$ 0.0053 &  -0.0071 $\pm$ 0.0046 & 0.0419 $\pm$ 0.0037  &  -0.0518 $\pm$ 0.0038 & 0.0405 $\pm$ 0.0103 \\ 
d006 & 192.3968 & -64.5231 & 19288 &  -0.0143 $\pm$ 0.0032 & 0.0261 $\pm$ 0.0025 &  0.0327 $\pm$ 0.0052 & 0.0740 $\pm$ 0.0059 &  -0.0114 $\pm$ 0.0049 & 0.0523 $\pm$ 0.0039  &  -0.0417 $\pm$ 0.0042 & -0.0021 $\pm$ 0.0108 \\ 
d007 & 195.7848 & -64.4954 & 22795 &  0.0225 $\pm$ 0.0026 & 0.0038 $\pm$ 0.0021 &  0.0603 $\pm$ 0.0045 & 0.0404 $\pm$ 0.0051 &  -0.0257 $\pm$ 0.0043 & 0.0629 $\pm$ 0.0034  &  -0.0564 $\pm$ 0.0035 & 0.0336 $\pm$ 0.0091 \\ 
d008 & 199.1573 & -64.3908 & 21183 &  0.0095 $\pm$ 0.0027 & 0.0237 $\pm$ 0.0023 &  0.0417 $\pm$ 0.0048 & 0.0539 $\pm$ 0.0057 &  -0.0145 $\pm$ 0.0043 & 0.0511 $\pm$ 0.0037  &  -0.0361 $\pm$ 0.0034 & -0.0164 $\pm$ 0.0100 \\
\end{longtable}
\end{landscape}
\normalsize



\section{Variation of the photometric coefficients per tile}
\begin{figure*}[h]
   \centering
  \includegraphics[width=13cm]{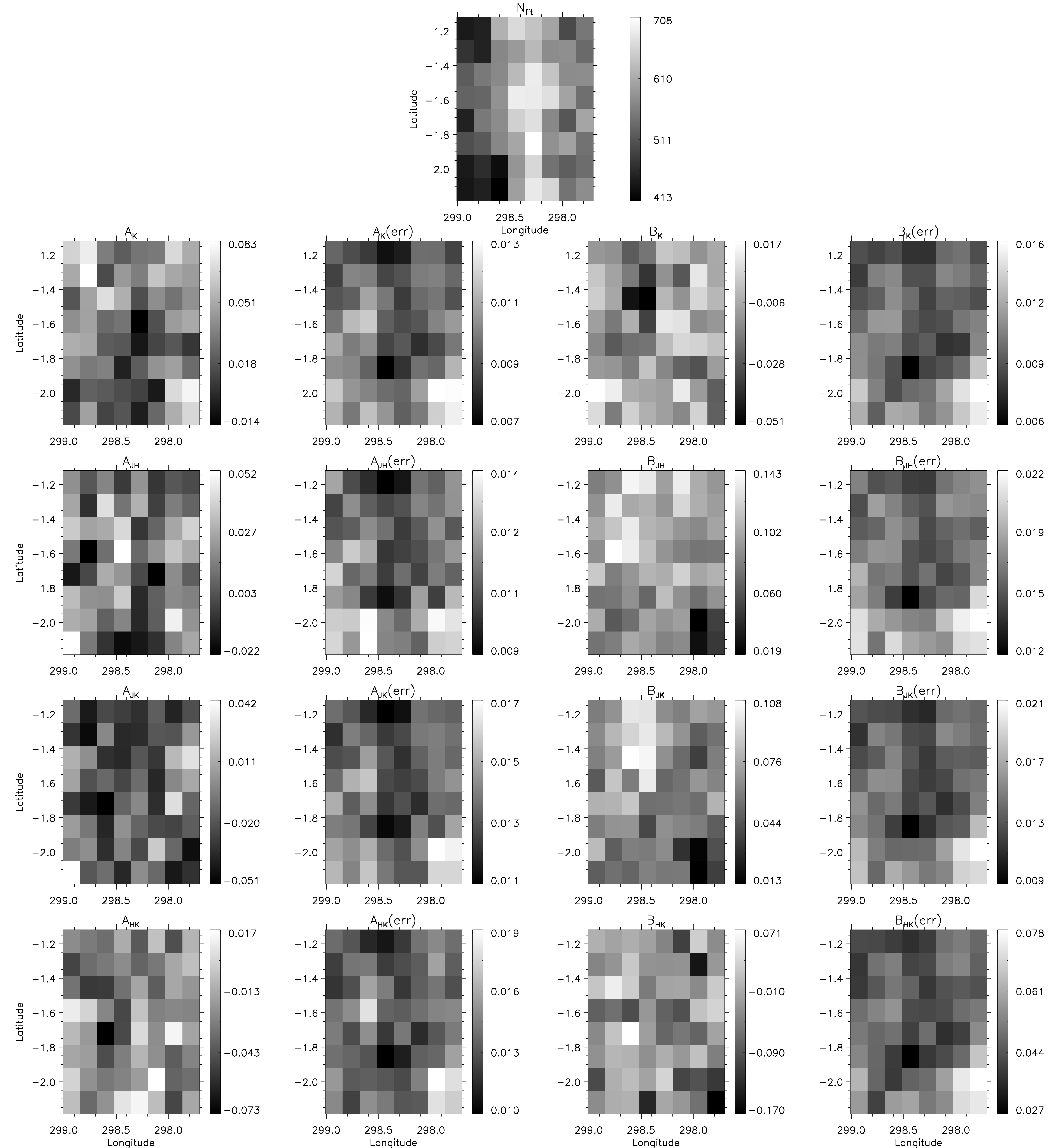}
   \caption{Variations of the coefficients for the photometric
     transformations across a single tile, tile $d003$. The complete 
 tile has been divided in 64 parts ($8\times8$). (\emph{Top}), number of 
 stars per sub-field. (\emph{Second row to bottom}), map of coefficients of
 the photometric transformations and their respective errors.
 }
              \label{fig:coeffsal_arr}%
    \end{figure*}

Thus far we presented photometric transformations calculated
for each tile.  Here we 
 explore the variation of the transformation coefficients across single tiles. We divided a
 complete tile in 64 (8$\times$8) parts.
 Each of these 64 sub-fields (1597$\times$1957 pixels) was used to calculate the transformation
 coefficients with the same procedure applied to complete tiles.
 We chose two tiles with different reddenings, namely tiles \emph{d003} and \emph{d050} (with modified SFD reddenings
 $E(B-V)\simeq 1.18$ and  $E(B-V)\simeq 3.29$, respectively).
  The average and dispersion, which for both fields are 
  shown in Table \ref{table:grid}, were weighted by the number of stars per
  sub-tile. Figure
  \ref{fig:coeffsal_arr} shows the map of the coefficients for tile
  \emph{d003}. 
  In general, there is general agreement between the average
  of the coefficients and those calculated for the complete field
  (Table \ref{table:tablecoef1}).   In some cases the variations are
  beyond the error bars, which may be attributed to population and extinction variations, or small statistics ($\sim 300$ stars) per sub-field.  
  Furthermore, the observed weighted dispersion 
  displays similar values for most of the coefficients $\sim 0.02\  mag$, 
  with the exception of $\beta_{HK}$, which exhibits a dispersion three
  times larger ($\sim 0.06\ mag$). Again, the observed dispersion in 
  $\beta_{HK}$ (Fig. \ref{fig:extcor}) appears sensitive to extinction.

\begin{table*}
\caption{Average and dispersion of transformation coefficients between
  2MASS and VVV photometric systems for 104 subfields in tiles d003 and d050 
  }             
\label{table:grid}      
\small
\centering                          
\begin{tabular}{l r l c r l}        
\hline\hline                 
& \multicolumn{2}{c}{$d003$} & & \multicolumn{2}{c}{$d050$} \\
\cline{2-3}
\cline{5-6}  
Coeff & $avg \ \ \ \ \ \ \ \ \ \ \ $ & $\ \ \ \ \ \ \ \ \  Stdev$   & & $avg \ \ \ \ \ \ \ \ \ \ \ $ & $\ \ \ \ \ \ \ \ \ Stdev$   \\    
\hline                        
$\alpha_K$    &   0.0331  $\pm$  0.0027  &  0.0221  $\pm$  0.0017  &  &     0.0423  $\pm$  0.0029  &  0.0237  $\pm$  0.0020 \\
$\beta_K$     &  -0.0103  $\pm$  0.0018  &  0.0140  $\pm$  0.0013  &  &    -0.0217  $\pm$  0.0026  &  0.0227  $\pm$  0.0037 \\
$\alpha_{JH}$ &   0.0135   $\pm$  0.0024  &  0.0180  $\pm$  0.0013  &  &     0.0511  $\pm$  0.0028  &  0.0217  $\pm$  0.0017  \\
$\beta_{JH}$  &  0.0948   $\pm$  0.0028  &  0.0238  $\pm$  0.0027   &  &     0.0789  $\pm$  0.0019  &  0.0153  $\pm$  0.0014  \\
$\alpha_{JK}$ & -0.0165   $\pm$  0.0024  &  0.0200  $\pm$  0.0016  &  &    -0.0241  $\pm$  0.0027  &  0.0217  $\pm$  0.0017  \\
$\beta_{JK}$  &  0.0640   $\pm$  0.0024  &  0.0195  $\pm$  0.0018  &  &     0.0680  $\pm$  0.0026  &  0.0206  $\pm$  0.0025  \\
$\alpha_{HK}$ & -0.0231   $\pm$  0.0023  &  0.0184  $\pm$  0.0016  &  &    -0.0708  $\pm$  0.0033  &  0.0270  $\pm$  0.0027  \\
$\beta_{HK}$  &  -0.0340  $\pm$  0.0061  &  0.0482  $\pm$  0.0042  &  &     0.0352  $\pm$  0.0067  &  0.0555  $\pm$  0.0060  \\

\hline                                   
\end{tabular}
\end{table*}

\vspace{4cm}

\section{Transformation coefficients divided by latitude and longitude}
\begin{landscape}
\centering
\begin{table}
\begin{minipage}{\linewidth}
\renewcommand{\footnoterule}{}
\caption{Fourth-order polynomial coefficients for the fitting of the
  photometric coefficients as a function of the Galactic longitude }             
\label{table:rscoef}      
\centering                          
\begin{tabular}{l r r r r r}        
\hline\hline                 
Coeff \footnote{Coefficients applied to a fitting equation of the form: $y= C_0 + C_1 x + C_2 x^2 + C_3 x^3 + C_4 x^4 $, where $x=l$, the Galactic longitude} &  $C_0$ \ \ \ \ \ \ \ \ \ \ \ \ \ & $C_1$ \ \ \ \ \ \ \ \ \ \ \  & $C_2$ \ \ \ \ \ \ \ \ \ \ \ \ \ \ \ \ \ \  \ \ & $C_3$ \ \ \ \ \ \ \ \ \ \ \ \ \ \ \ \ \ \ & $C_4$  \ \ \ \ \ \ \ \ \ \ \ \ \ \ \ \ \ \ \\
\hline                        
$\alpha_K$  ($|b| < 1^{\circ}$)                     &   -365.4430 $\pm$  34.3853 &  4.4934 $\pm$   0.4285 &-2.0710e-02 $\pm$ 0.2000e-02 & 4.2407e-05 $\pm$ 0.4147e-05 &-3.2551e-08 $\pm$ 0.3220e-08 \\  
$\beta_K$                                                &      18.8419 $\pm$  27.2586 & -0.1829 $\pm$   0.3391 & 0.6150e-03 $\pm$ 1.5809e-03 &-0.78657e-06 $\pm$ 3.2722e-06 & 2.3628e-10 $\pm$ 0.2538e-09 \\  
$\alpha_{JH}$                                            &    -417.7458 $\pm$  49.0418 &  5.2539 $\pm$   0.6116 &-2.4749e-02 $\pm$ 0.2858e-02 & 5.1752e-05 $\pm$ 0.5930e-05 &-4.0528e-08 $\pm$ 0.4609e-08 \\  
$\beta_{JH}$                                             &     317.8541 $\pm$  57.8718 & -4.0524 $\pm$   0.7205 & 1.9351e-02 $\pm$ 0.3360e-02 &-4.1006e-05 $\pm$ 0.6960e-05 & 3.2534e-08 $\pm$ 0.5400e-08 \\  
$\alpha_{JK}$                                            &     -41.0102 $\pm$  52.1432 &  0.5933 $\pm$   0.6501 &-3.1379e-03 $\pm$ 3.0363e-03 & 7.2314e-06 $\pm$ 6.2971e-06 &-6.1507e-09 $\pm$ 4.8930e-09 \\  
$\beta_{JK}$                                             &     424.2013 $\pm$  42.5080 & -5.3723 $\pm$   0.5290 & 2.5478e-02 $\pm$ 0.2467e-02 &-5.3618e-05 $\pm$ 0.5107e-05 & 4.2249e-08 $\pm$ 0.3962e-08 \\  
$\alpha_{HK}$                                            &     843.9939 $\pm$  50.2942 &-10.4583 $\pm$   0.6269 & 4.8559e-02 $\pm$ 0.2928e-02 &-1.0013e-04 $\pm$ 0.0607e-04 & 7.7367e-08 $\pm$ 0.4718e-08 \\  
$\beta_{HK}$                                             &    -442.9768 $\pm$ 129.7205 &  5.3851 $\pm$   1.6147 &-2.4535e-02 $\pm$ 0.7530e-02 & 4.9648e-05 $\pm$ 1.5594e-05 &-3.7649e-08 $\pm$ 1.2099e-08 \\  
 & & & & &  \\
\hline
$\alpha_K$  ($-2.1^{\circ} \geq |b| \geq 1^{\circ}$) & -1923.7242 $\pm$  37.1316 & 24.1147 $\pm$   0.4637 &-1.1324e-01 $\pm$ 0.0217e-01 & 2.3610e-04 $\pm$ 0.0451e-04 &-1.8439e-07 $\pm$ 0.0351e-07 \\   
$\beta_K$                                                &      15.0531 $\pm$  25.7626 & -0.1462 $\pm$   0.3211 & 0.4969e-03 $\pm$ 1.4994e-03 &-0.6634e-06 $\pm$ 3.1094e-06 & 0.2439e-09 $\pm$ 2.4160e-09 \\  
$\alpha_{JH}$                                           &    -448.6523 $\pm$  55.9766 &  5.5559 $\pm$   0.6994 &-2.5813e-02 $\pm$ 0.3274e-02 & 5.3338e-05 $\pm$ 0.6807e-05 &-4.1360e-08 $\pm$ 0.5302e-08 \\  
$\beta_{JH}$                                             &     588.0175 $\pm$  59.3255 & -7.3890 $\pm$   0.7397 & 3.4798e-02 $\pm$ 0.3456e-02 &-7.2778e-05 $\pm$ 0.7170e-05 & 5.7031e-08 $\pm$ 0.5573e-08 \\  
$\alpha_{JK}$                                            &    1402.2977 $\pm$  59.4223 &-17.4729 $\pm$   0.7424 & 8.1554e-02 $\pm$ 0.3475e-02 &-1.6899e-04 $\pm$ 0.0722e-04 & 1.3115e-07 $\pm$ 0.0563e-07 \\  
$\beta_{JK}$                                             &     250.0838 $\pm$  42.4663 & -3.1881 $\pm$   0.5294 & 1.5220e-02 $\pm$ 0.2473e-02 &-3.2238e-05 $\pm$ 0.5130e-05 & 2.5564e-08 $\pm$ 0.3987e-08 \\  
$\alpha_{HK}$                                            &    1462.7856 $\pm$  56.5467 &-18.1479 $\pm$   0.7064 & 8.4378e-02 $\pm$ 0.3306e-02 &-1.7426e-04 $\pm$ 0.0687e-04 & 1.3487e-07 $\pm$ 0.0535e-07 \\  
$\beta_{HK}$                                             &   -1269.9341 $\pm$ 122.9108 & 15.6974 $\pm$   1.5323 &-7.2724e-02 $\pm$ 0.7158e-02 & 1.4966e-04 $\pm$ 0.1485e-04 &-1.1543e-07 $\pm$ 0.1154e-07 \\  
\hline                                   
\end{tabular}
\end{minipage}
\end{table}

\centering
\begin{table}
\begin{minipage}{\linewidth}
\renewcommand{\footnoterule}{}
\caption{Third-order polynomial coefficients for the fitting of the
  photometric coefficients as a function of the Galactic latitude }             
\label{table:rscoef}      
\centering                          
\begin{tabular}{l r r r r}        
\hline\hline                 
Coeff \footnote{Coefficients applied to a fitting equation of the form: $y= C_0 + C_1 x + C_2 x^2 + C_3 x^3 $, where $x=b$, the Galactic latitude} &  $C_0$ \ \ \ \ \ \ \ \ \ \ \ \ \ & $C_1$ \ \ \ \ \ \ \ \ \ \ \  & $C_2$ \ \ \ \ \ \ \ \ \ \ \ \ \ \ \ \ \ \  \ \ & $C_3$ \ \ \ \ \ \ \ \ \ \ \ \ \ \ \ \ \ \  \\
\hline                        
$\alpha_K$   ($l\geq 320^{\circ}$)      &      8.7137e-02 $\pm$ 0.0248e-02 & 3.2361e-02 $\pm$ 0.0453e-02 &-1.8083e-02 $\pm$ 0.0114e-02 &-1.4078e-02 $\pm$ 0.0173e-02 \\ 
$\beta_K$         &     -1.5320e-02 $\pm$ 0.0156e-02 &-2.3980e-03 $\pm$ 0.2848e-03 & 1.6483e-03 $\pm$ 0.0769e-03 & 1.4568e-03 $\pm$ 0.1106e-03 \\  
$\alpha_{JH}$      &     2.4836e-02 $\pm$ 0.0396e-02 &-1.8102e-02 $\pm$ 0.0723e-02 & 9.2059e-04 $\pm$ 1.7754e-04 & 7.8566e-03 $\pm$ 0.2762e-03 \\ 
$\beta_{JH}$       &     6.5333e-02 $\pm$ 0.0386e-02 &-3.1511e-03 $\pm$ 0.7048e-03 & 2.8729e-03 $\pm$ 0.1817e-03 & 5.4973e-04 $\pm$ 2.7140e-04 \\  
$\alpha_{JK}$      &    -5.6450e-02 $\pm$ 0.0415e-02 &-2.5840e-02 $\pm$ 0.0756e-02 & 1.6496e-02 $\pm$ 0.0185e-02 & 1.1324e-02 $\pm$ 0.0289e-02 \\ 
$\beta_{JK}$       &     5.4228e-02 $\pm$ 0.0269e-02 &-1.8258e-03 $\pm$ 0.4919e-03 &-7.0817e-04 $\pm$ 1.2830e-04 &-2.8354e-04 $\pm$ 1.8979e-04 \\  
$\alpha_{HK}$      &    -6.9442e-02 $\pm$ 0.0393e-02 &-1.5938e-02 $\pm$ 0.0717e-02 & 1.3718e-02 $\pm$ 0.0177e-02 & 5.8553e-03 $\pm$ 0.2739e-03 \\ 
$\beta_{HK}$       &     2.2701e-02 $\pm$ 0.0754e-02 & 6.9031e-03 $\pm$ 1.3767e-03 &-1.2861e-02 $\pm$ 0.0374e-02 &-2.4821e-03 $\pm$ 0.5351e-03 \\ 
 & & & &  \\
\hline                                   
$\alpha_K$  ($l < 320^{\circ}$)      &      3.8479e-02 $\pm$ 0.0196e-02 &-9.3109e-03 $\pm$ 0.3560e-03 &-1.9789e-03 $\pm$ 0.1088e-03 & 1.9851e-03 $\pm$ 0.1415e-03 \\ 
$\beta_K$         &     -1.3918e-02 $\pm$ 0.0141e-02 & 4.5106e-03 $\pm$ 0.2552e-03 & 1.9777e-03 $\pm$ 0.0875e-03 &-1.3459e-03 $\pm$ 0.1043e-03 \\
$\alpha_{JH}$      &     3.6107e-02 $\pm$ 0.0293e-02 &-4.8033e-03 $\pm$ 0.5342e-03 &-5.9765e-03 $\pm$ 0.1543e-03 & 3.2205e-03 $\pm$ 0.2098e-03 \\ 
$\beta_{JH}$       &     7.4869e-02 $\pm$ 0.0335e-02 &-4.9043e-03 $\pm$ 0.6098e-03 & 2.9709e-03 $\pm$ 0.1888e-03 & 3.3755e-04 $\pm$ 2.4307e-04 \\
$\alpha_{JK}$      &    -2.4255e-02 $\pm$ 0.0316e-02 & 6.5936e-03 $\pm$ 0.5748e-03 & 2.8184e-03 $\pm$ 0.1670e-03 &-5.2040e-04 $\pm$ 2.2604e-04 \\ 
$\beta_{JK}$       &     5.6893e-02 $\pm$ 0.0240e-02 &-5.2755e-03 $\pm$ 0.4366e-03 &-6.8883e-04 $\pm$ 1.3891e-04 & 5.3354e-04 $\pm$ 1.7513e-04 \\
$\alpha_{HK}$      &    -5.6496e-02 $\pm$ 0.0299e-02 & 1.1284e-02 $\pm$ 0.0546e-02 & 9.4227e-03 $\pm$ 0.1601e-03 &-4.1742e-03 $\pm$ 0.2150e-03 \\ 
$\beta_{HK}$       &     1.4171e-02 $\pm$ 0.0665e-02 &-7.1747e-03 $\pm$ 1.2078e-03 &-1.4146e-02 $\pm$ 0.0409e-02 & 3.2066e-03 $\pm$ 0.4921e-03 \\ 
\hline
\end{tabular}
\end{minipage}
\end{table}

\end{landscape}

\end{appendix}

\end{document}